\pdfoutput=1

\documentclass{aa}

\usepackage{graphicx}
\usepackage{txfonts}
\usepackage[bookmarksopen=false,bookmarksnumbered=true,breaklinks=true, colorlinks=true,linkcolor=blue,citecolor=blue]{hyperref}

\newcommand{\modif}{}
\newcommand{\mmodif}{}

\begin{document}

   \title{Direct exoplanet detection and characterization using the ANDROMEDA method: Performance on VLT/NaCo data\thanks{Based on observations made with ESO Telescopes at the Paranal Observatory under programme ID 184.C-0567(D) and 60.A-9800(J).}}
   \titlerunning{Exoplanet imaging: Performance of the ANDROMEDA pipeline on VLT/NaCo data.}

   \author{
        F.~Cantalloube\inst{1,2,3}
        \and D.~Mouillet\inst{1,3}
        \and L.~M.~Mugnier\inst{2}
        \and J.~Milli\inst{1,3,4}
        \and O.~Absil\inst{5}\fnmsep\thanks{F.R.S.-FNRS Research Associate}
        \modif{\and C.~A.~Gomez~Gonzalez\inst{5}}
        \and G.~Chauvin\inst{1,3}
        \and J.-L.~Beuzit\inst{1,3}
        \and A.~Cornia\inst{2}
          }

   \institute{Univ. Grenoble Alpes, IPAG, F-38000 Grenoble, France\\
                \email{faustine.cantalloube@obs.ujf-grenoble.fr}
                \and Office National d'Etudes et de Recherches A\'erospatiales (ONERA),
Optics Department, BP 72, 92322 Ch\^atillon, France
                \and CNRS, IPAG, F-38000 Grenoble, France
                \and European Southern Observatory (ESO), Alonso de C\`ordova 3107, Vitacura, Casilla 19001, Santiago, Chile
                \and D\'epartement d'Astrophysique, G\'eophysique et Oc\'eanographie, Universit\'e de Li\`ege, All\'ee du Six Ao\^{u}t 17, 4000 Li\`ege, Belgium
             }

   \date{Received December 22, 2014; accepted August 04, 2015}

  \abstract
{
The direct detection of exoplanets with high-contrast imaging requires advanced data processing methods to disentangle potential planetary signals from bright quasi-static speckles. Among \modif{them}, angular differential imaging (ADI) permits  potential planetary signals with a known rotation rate to be separated  from instrumental speckles that are either statics or slowly variable. The method presented in this paper, called ANDROMEDA for \emph{ANgular Differential OptiMal Exoplanet Detection Algorithm} is based on a maximum likelihood approach to ADI and is used to estimate the position and the flux of any point source present in the field of view.
}
{
In order to optimize and  \modif{experimentally} validate this previously proposed method, we applied ANDROMEDA to real VLT/NaCo data. In addition to its pure detection capability, we investigated the possibility of defining simple and efficient criteria for automatic point source extraction able to support the \modif{processing} of large surveys.
}
{
To assess the performance of the method, we applied ANDROMEDA on VLT/NaCo data of TYC-8979-1683-1 which is surrounded by numerous bright stars and on which we added \modif{synthetic} planets of known position and flux in the field. In order to accommodate the real data properties, it was necessary to develop additional pre-processing and post-processing steps to the initially proposed algorithm. We then investigate\modif{d} its skill in the challenging case of a well-known target, $\beta$ Pictoris, whose companion is close to the detection limit and we  compared our results to those obtained by another method based on principal component analysis (PCA).
}
{
Application on VLT/NaCo data demonstrates the ability of ANDROMEDA to automatically detect and characterize point sources present in the image field.  We end up with a robust method bringing consistent results with a sensitivity similar to the \modif{recently} published algorithms, with only two parameters to be fine tuned. Moreover, the companion flux estimates are not biased by the algorithm parameters and do not require a posteriori corrections.
}
{
ANDROMEDA is an attractive alternative to current standard image processing methods that can be readily applied to on-sky data.
}


   \maketitle


\section{Introduction}

Among the thousands of exoplanets known today, less than $2\%$ have been detected by direct imaging. The methods most often used at the present date are  indirect, such as Doppler spectroscopy \citep{Marcy1993,Walker1995}, transit photometry \citep{Rosenblatt1971,Borucki1985}, or microlensing \citep{Cassan2012}, that have permitted  the population of inner planets (semi-major axis $\lesssim 5 \; AU$) of sub-Jovian mass ($\lesssim 2.10^{30} \; g$) to be characterized \citep{UdrySantos07,Schneider11}.
Direct imaging is motivated by its ability to detect massive young exoplanets on wider orbits and also to obtain spectra that provide information about their atmospheric composition. Thus, direct imaging, complementary with other indirect techniques, increases our statistics and understanding of exoplanets.

Very few exoplanets have been imaged up to the present day, and this is essentially due to the difficulty in achieving high contrast at small angular separation, as most giant exoplanets lie within 1\arcsec\  from their host star with a flux ratio of about $10^{5}$ to $10^{8}$ (depending on mass and age). In order to obtain a high resolution, observation with  very large ground-based telescopes is needed,  but the resolution is then limited by  atmospheric turbulence. This effect can be overcome by the use of an adaptive optics (AO) system, which performs a real-time correction of the incoming distorted wavefront. The second step is to \modif{handle} the very high contrast \modif{that exists} between the planet and its host star. \modif{The use of a coronagraph greatly helps by removing a large part of the coherent light from the star while preserving its close environment}. However, the optics are not perfect, \modif{and in long exposures the remaining aberrations are responsible for speckles in the image that} vary from less than minutes to hours \citep{Hinkley2007}; \modif{the speckles are} too slow to be averaged \modif{into a smooth} halo and too fast to be calibrated and removed a posteriori. These quasi-static speckles make exoplanet detection confusing since they are of the same angular size as the expected planetary signal and are often brighter. Consequently, at this stage it is necessary to apply elaborated image processing methods to disentangle the signal of an exoplanet from the remaining speckle noise.

\modif{The image processing methods that are the most widely used by astronomers today are based on the angular differential imaging (ADI) concept \citep{Marois06}. This technique relies on observations made in pupil tracking mode that fixes the pupil while the image rotates with time. Several algorithms are used to build a \emph{reference} point spread function (PSF) that should reproduce as accurately as possible the speckle pattern to be subtracted, but not the rotating signature of a potential astronomical source around the on-axis star. 
Among them, Classical ADI (cADI) uses a median image to build the reference PSF, and LOCI (Locally Optimized Combination of Images) \citep{Lafreniere07} 
uses a linear combination of images to build it and locally optimizes the speckle subtraction. The principal component analysis (PCA), in the form of PynPoint \citep{Amara2012} or KLIP \citep{Soummer2012}, uses an expansion of eigenimages to build the reference PSF. Numerous variations of these methods are studied and are used to improve the speckle subtraction.}

The  ANgular DiffeRential OptiMal Exoplanet Detection Algorithm (ANDROMEDA) method\modif{presented} in this paper is an advanced way of exploiting the ADI technique in the framework of inverse problems. ANDROMEDA uses a maximum likelihood estimation to detect planetary signals and retrieve the projected distance between the planet and the star, as well as the contrast between the two components. From these estimations, and knowing the age of the host star, sophisticated models (e.g., \cite{Allard2012,Spiegel2012,Marleau2014} for young stars) provide estimations of the companions' mass and surface temperature. These results are useful to constrain the current models of star and planetary formation \modif{and their evolution}.

In this paper we first present and explain in detail the different steps performed by the algorithm ANDROMEDA to properly process real images and we discuss the hypothesis made to check whether it is consistent for real data (Sect.~\ref{sect-method}).
We then explain how to use the output provided by ANDROMEDA to automatically detect the point sources in the images and to extract both their flux and position (Sect.~\ref{sect-Toolbox}).
The next section is an application of the full procedure to real VLT/NaCo data, which consists of a bright star (TYC-8979-1683-1) surrounded by numerous background stars acting as point sources to be detected, and in which we injected synthetic companions in order to assess the performance of ANDROMEDA (Sect.~\ref{sect-NaCo}).
The following section shows results using VLT/NaCo data of the well-known case of Beta-Pictoris, imaged under fair conditions and using an AGPM coronagraph \citep{Mawet2005}, thus approaching the quality of future high-resolution and high-contrast imaging systems \citep{Absil2013}. This last part (Sect.~\ref{sect-betapic})   compares the performance obtained by using ANDROMEDA with another commonly used method, the PCA.


\section{ANDROMEDA's principle}
\label{sect-method}

ANDROMEDA uses a statistical approach to discriminate planetary signals from the remaining speckles. \modif{The first step is to perform ADI in order to create \emph{\emph{differential images}} in which speckles that are stable enough to be distinguished from a rotating companion signature are removed. If a companion is present, a specific signature appears at its location that we can model according to its flux and initial position parameters. The second and original step of the method is to search for such a signature in the differential images by using a maximum likelihood estimation (MLE) of its position and intensity. This estimation is made from a consistent model of the  data set, given a statistical model of its noise distribution as first presented in \cite{Mugnier09}, it is similar to the approach proposed for the processing of DARWIN/TPF-I data in~\cite{Thiebaut} and to the works by \cite{Smith09} for a perfectly fixed speckle field.}

\modif{The purpose of this paper is to apply ANDROMEDA to real data from the VLT/NaCo instrument \citep{NACO1, NACO2}. This implied  adapting the method described in \cite{Mugnier09}.} The different processing steps performed by the upgraded version of ANDROMEDA presented and used in this paper are the following. Section~\ref{pre-processing} describes how \modif{and why} the artifacts with low spatial frequency  are removed from the \modif{raw} images in a pre-processing step. \modif{Section~\ref{Sec-DI} describes the angular image difference imaging that is then} performed in order to suppress the vast majority of the starlight from the images.
Section~\ref{pseudo-data} \modif{defines} the mathematical model \modif{corresponding to the resulting differential images} along with their dependency with the sought position\modif{s} and amplitude\modif{s} of the potential companions.
\modif{Section~\ref{max-likelihood} explains how the problem is inverted between the differential image and its model by using a maximum likelihood which analytically finds an estimation of both the position and flux of the potential companions.}
\modif{Finally,  Sect.~\ref{post-processing} describes the simple and effective post-processing that is performed on the results in order to compensate for the deviation from the noise model (assumed to be white and Gaussian in the differential images) that appears in practice on real data.}

\subsection{Pre-processing: Filtering out low-frequency artifacts}
\label{pre-processing}

In both the raw and the reduced images there are some strong inhomogeneities of low spatial frequency that are not stable \modif{and thus} disturb  planetary detection. These large-scale structures in the images, which vary slowly from one frame to another, might originate from temporal variation of the residual turbulence. As they cannot be fully subtracted via the ADI process and they are not included in the \modif{image} model used to \modif{perform the MLE}, these low frequencies must be removed.

\begin{figure}[!h]
\centering
\resizebox{\hsize}{!}{\includegraphics{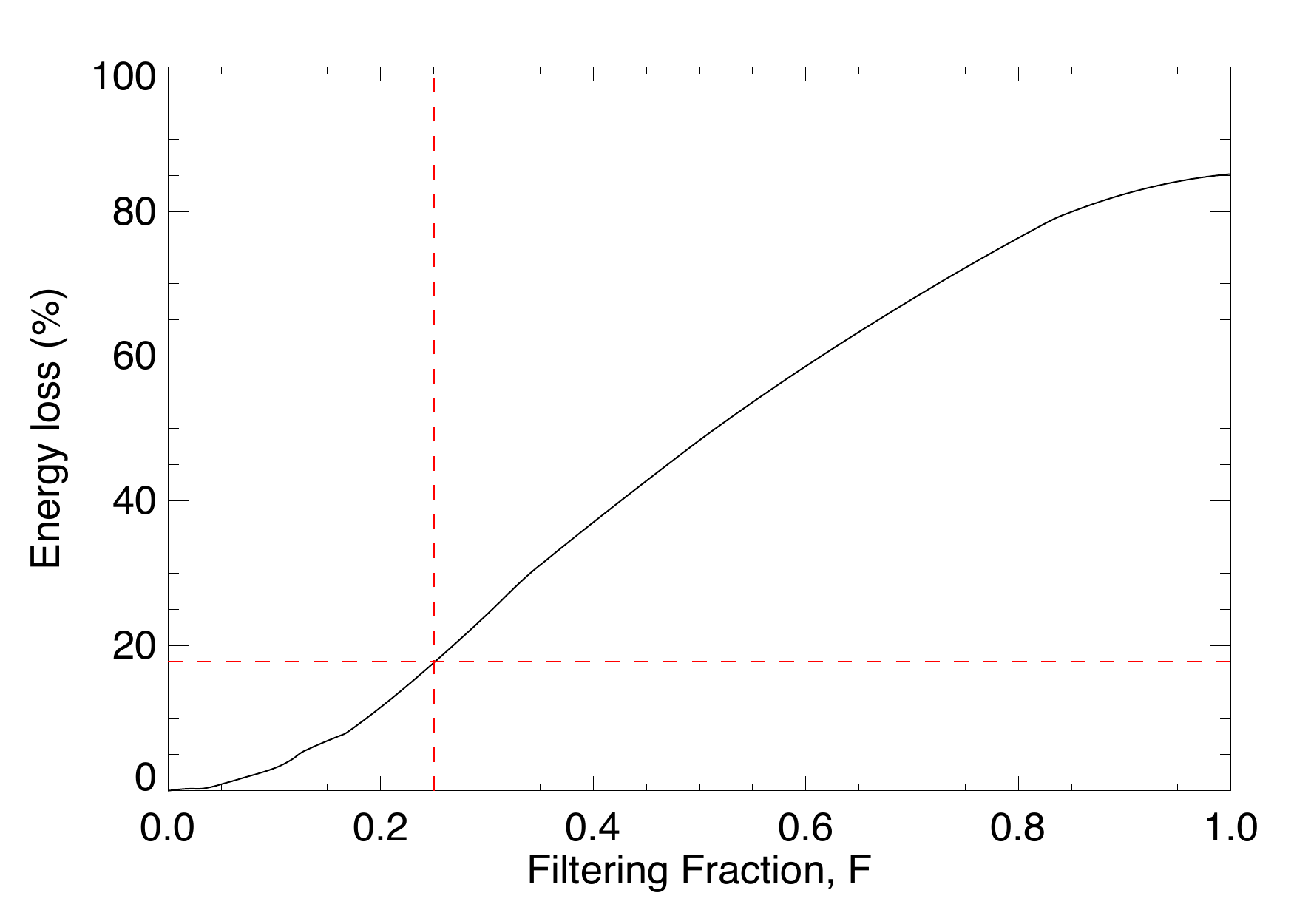}}
\caption{\modif{Energy} loss as a function of the filtering fraction when filtering a simple Airy pattern (sampled at \modif{the Nyquist rate}) with a high-pass Hanning Fourier filter \modif{(solid line). Energy loss for a usual filtering fraction of 1/4 is shown with the red dashed line.} }
\label{fig-filteringLoss}
\end{figure}

A pre-processing of the data has thus been introduced in ANDROMEDA to make the detection of point sources easier within the stellar halo. This pre-processing consists in eliminating these disturbing structures by \modif{applying a high-pass filter in the Fourier space. Half of a Hann function has been chosen to build this filter's profile in order to avoid Gibbs effects, due to discontinuity, that appear when going from the Fourier to the real space. This filter} attenuates the spatial frequencies lower than a chosen cutoff frequency defined as $f_c = F \cdot f_N$, where $f_N$ is the Nyquist frequency and $F$ is a factor in the range $0$-$1$ (hereafter called the filtering fraction). The parameter $F$ is  user-defined and  must be chosen carefully to  efficiently remove the low-frequency artifacts while preserving \modif{most of the energy of the point source signals} in the images. In order to quantify \modif{the signal loss due to this filtering}, we took an Airy pattern and applied the filtering to it. The \modif{energy} loss is then quantified as follows:
\begin{eqnarray}
 \mmodif{E}_{loss} = 1 - \frac{\int \mmodif{{(airy_{filtered})}^2}}{\int \mmodif{{(airy_{non-filtered})}^2}}
.\end{eqnarray}
The obtained results are gathered in a graph showing the \modif{energy} loss as a function of the filtering fraction (Fig.~\ref{fig-filteringLoss}).

Comparisons and tests on simulation and real data show that even though this procedure slightly reduces the intensity of the planetary signal on individual frames, the filtering is worthwhile for planet detection \modif{and does not introduce any error on the estimated flux.} \modif{In practice, a filtering fraction of about one fourth is recommended when dealing with VLT/NaCo data, leading to an energy loss of about 18\% (see Fig.~\ref{fig-filteringLoss}).} 

\subsection{Construction of differential images}
\label{Sec-DI}
\subsubsection{Angular differential imaging (ADI)}
\label{ADI}
The  ADI technique consists in taking advantage of the alt-azimuthal mount of the telescope in use: the pupil and the field  rotate  at different deterministic velocities during the observation.
In order to perform ADI, one can choose to fix the pupil (by using a pupil derotator if needed) to stabilize the aberrations in the image, while the observed field rotates as the parallactic angle. Therefore, these images can be later subtracted from each other in an optimized way to reduce as much as possible the speckle noise in the resulting \modif{differential image}. This method was first introduced by \cite{Marois06} and is the basis of most image processing methods dedicated to exoplanet detection today.

In order to choose the two images to be subtracted, a compromise has to be made so that (i) the turbulence strength and the quasi-static aberrations do not have time to evolve significantly between the two images, and (ii) the object of interest has rotated enough in the field between the two images in order not to cancel its own signal during the subtraction. Assuming that the most correlated data are the closest in time, we chose two images that are as close as possible in time but within a limitation to avoid the self-subtraction of the companion. This limitation is the minimum distance between the position of the planetary signal in the two images ($\delta_{\rm min}$), which is a key user-defined parameter whose value must be adjusted according to the speckle lifetime and instrumental stability \modif{(typically  $1.0 \lambda/D$, see Table~\ref{tab-parameters})}. \modif{As the displacement between two images is dependent upon} the distance between the star and the object of interest, the ADI is performed over concentric annuli of constant width $dr$ \modif{to accommodate this dependency}, which is another user-defined parameter \modif{(typically  $1.0 \lambda/D$, see Table~\ref{tab-parameters}).}
For each of these annuli, \modif{each image of the cube is subtracted from the first following image (or previous image if needed)  that meets condition (ii). If a couple is found, the subtraction results in the creation of a \emph{differential image}, indexed $k$, where $k \in [1;N_k]$, with $N_k$ the total number of couples found within the cube at the regarded distance from the star. If for a given frame, no other frame satisfies  condition (ii), no pair can be built with this frame which is therefore not used. Thus $N_k$ is less than or equal to the total number of images in the cube.}

As the two images to be subtracted are likely not to  have the same intensity distribution, one of the image needs to be adjusted with respect to the other via \modif{a} scaling factor, noted $\gamma_k$. These intensity variations are primarily caused by variations of the seeing in the course of the observation. Each \modif{differential image} $\Delta({\bf r},k)$ resulting from the subtraction of a couple $k$ of \modif{images} $i({\bf r}, t)$ is then of the form
\begin{eqnarray}
\Delta({\bf r},k) = i_{\mmodif{k}}({\bf r}, \mmodif{t_1}) - \gamma_k \, i_{\mmodif{k}}({\bf r}, \mmodif{t_2}) \;.
\label{eqADI}
\end{eqnarray}
It can be seen in both the simulations and the real data that the flux difference depends on the distance to the star. As a consequence, the scaling factor is computed for each couple of annuli, over an \emph{optimization area} that is slightly wider than the effective \emph{subtraction area} in order to avoid discontinuities between adjacent annuli \citep{Cornia10}. The ratio optimization to subtraction area ($R_A$) is a third user-defined parameter, set constant for all annuli \modif{(typically  $2.0$, see Table~\ref{tab-parameters})}.

Several considerations guided us in the design of the best computation of this scaling factor.
\modif{Because the high-pass filtered images have a zero mean, it is not possible to compute the scaling factor $\gamma_k$ with a simple ratio of the total intensities in each image. Instead the scaling factor is better estimated by a least-squares fit that minimizes $\lVert i_{\mmodif{k(t_1)}}(\mathbf{r}) - \gamma_k \;  i_{\mmodif{k(t_2)}}(\mathbf{r}) \rVert^2$ \citep{Cornia10}. \modif{In the following, we refer to this method as the \emph{LS optimization}}. We note that optimization depending on position in the field has been implemented in other methods of exoplanetary detection, as in \cite{Marois06,Lafreniere07}.} \modif{Another way to subtract the images is also presented here, after taking two aspects} into consideration: first the PSF may not vary linearly and if there is a sudden evolution in the turbulence strength, \modif{the image can be better fitted by an affine law. Thus, the differential image can be constructed by}
\begin{eqnarray}
\Delta(\mathbf{r},k) = i_{\mmodif{k(t_2)}}(\mathbf{r}) - \gamma_k  \, i_{\mmodif{k(t_1)}}(\mathbf{r}) - \gamma'_k
\label{affine}
.\end{eqnarray}
Second, the least-squares  is \modif{sensitive} to outliers and if, \modif{from one image to the other, a speckle or a planetary signal intensity has  varied significantly, their high signal} is taken into account to compute the scaling factor, which results in a flux offset in the differential image. In order to avoid the latter bias, we additionally implemented a robust fit, using a $L_1$ norm \modif{that can be chosen} instead of the $L_2$ norm. The scaling factors $\gamma_k$ and $\gamma'_k$ are then computed by minimizing: $\|i_{\mmodif{k}}(\mathbf{r}) - \gamma_k  \,.\, i_{\mmodif{k}}(\mathbf{r}) - \gamma'_k \|_{L_1}$. \modif{In case there are such outliers corresponding to high flux and variable signals (either speckles or planetary signals), this way of combining the image pair -- hereafter called \emph{L1-affine optimization} -- has proven equally or more efficient  at reducing the speckle noise on both simulated and real data, while preserving the planetary signal, if any. This L1-Affine optimization is particularly good at very close separation where the intensity of the numerous speckles varies a lot from one image of the couple to the other due to the poor rotation field velocity.}

After this operation, the \modif{differential images} $\mmodif{\Delta(\mathbf{r},k)}$ are obtained \modif{for every distance to the star}.
\modif{We note that even if the image couples are determined for specific annuli and  the optimizations are computed over annular areas including these annuli, the whole images are actually subtracted one from another to build the differential images.} 
By minimizing the residuals in these data, the noise has been partly whitened \modif{spatially} \citep{Cornia10, Mawet2014}. Since ANDROMEDA relies on solving an inverse problem, it is also useful to estimate the variance map $\sigma_\Delta^2(\mathbf{r})$ of the \modif{differential images}. Currently it is \modif{estimated} empirically over the time dimension $k$ as a function of the position $\mathbf{r}$:
\begin{eqnarray}
\sigma_\Delta^2(\mathbf{r}) = \langle [\Delta(\mathbf{r},k)]^2 \rangle_k - [\langle \Delta(\mathbf{r},k) \rangle_k]^2
\label{eq-variance}
.\end{eqnarray}
\modif{We note that when the observer has other independent ways to estimate the variance map, in particular taking  the temporal variations into account, it could be introduced here. In the following, it is nevertheless assumed that the noise variance is stationary in time and the definition of the variance map given above is used, which is satisfactory for our study.}

\subsubsection{Other possible combinations of images}
\label{othersDI}
\modif{Because ANDROMEDA was  developed with the objective of being used to process large amount of survey data from second-generation instruments, it was  proposed with a simple ADI subtraction to whiten the noise. This allows the algorithm to run quickly (in particular, it is possible to parallelize the computation of each operation), and to have a very low number of user-parameters to be adjusted. However, many other approaches can be considered to build up differential images that could also be  included in the ANDROMEDA scheme, as long as the impact on the potential companion signature can be evaluated and inserted in the model. For instance, it would be possible to merge this method with other subtraction algorithms, such as LOCI or PCA, that are known  to be able to reduce  the speckle noise very efficiently.}

\modif{Other techniques can be used to whiten the noise: instead of exploiting only the temporal information, ANDROMEDA could deal with spectral or polarization redundancy, for instance, and perform, respectively, spectral differential imaging (SDI) \citep{Racine99,Marois00} or polarimetric differential imaging (PDI) \citep{Kuhn01-PDI} in addition to the ADI. An optional SDI for ANDROMEDA has been implemented and discussed in \cite{Cornia10} and simple tests have been led on images taken either with a dual band imager (DBI) or an integral field spectrograph (IFS), such as IRDIS or IFS on the VLT/SPHERE instrument, but this work is beyond the scope of this paper.}

\subsection{Building a model of the \modif{differential images}}
\label{pseudo-data}
\begin{figure}[!h]
\centering
\resizebox{\hsize}{!}{\includegraphics{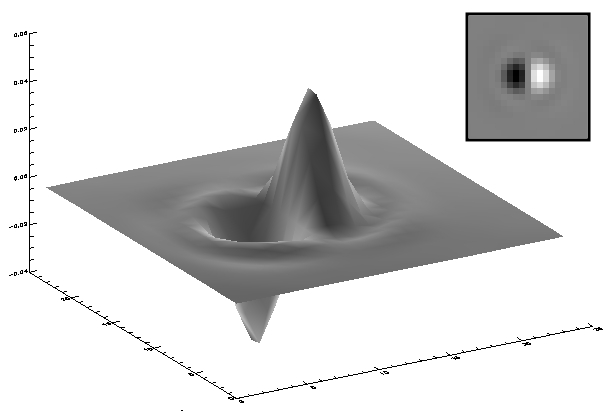}
                      \includegraphics{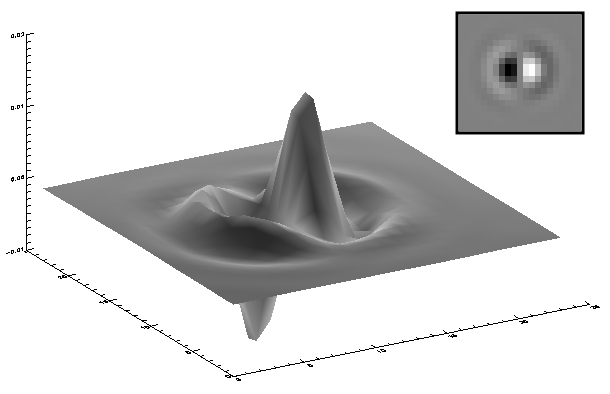}}
\caption{Illustration of the planet signatures obtained when performing the ADI, setting $\delta_{min}=0.5 \lambda /D$. These have been obtained by simulating  two identical noiseless planetary PSF, separated by $\delta_{min}$ and then subtracted one from the  other.
\textit{Left:} Planet signature obtained without high-pass filtering of the raw data. 
\textit{Right:} Planet signature obtained when the raw data have been filtered of their low frequencies \modif{($\mmodif{F=1/4}$)}. This time secondary opposite lobes  surround the main lobe.}
\label{fig-PlanetSignature}
\end{figure}

Once the \modif{differential images} have been created,  evidence of the presence of a planetary signal in the original image is sought inside all of these \modif{differential images}.
After performing the ADI, if a companion is indeed present in the field of view, a peculiar signature \modif{appears} in the \modif{differential images}. This so-called planet signature is naturally the difference of two planetary signals separated by at least $\mmodif{\delta_{\rm min}}$. Figure~\ref{fig-PlanetSignature} illustrates the shape of the expected planet signature obtained either with or without high-pass filtering of the raw data.
In order to properly estimate the position and flux of a potential exoplanet using a maximum likelihood approach, we now define a model for these \modif{differential images}.
Assuming that the stellar halo is fully suppressed by the subtractions and that a planet is present in the field of view, each \modif{differential image} can be modeled as
\begin{eqnarray}
\Delta({\bf r},k) =  a \cdot {\bf p}({\bf r},k;{\bf r}_0) + {\bf n}({\bf r},k)\;,
\label{eq1bis}
\end{eqnarray}
where the planet's flux $a$ and the position of the planet on the first image of the sequence (initial position) ${\bf r}_0$ are the two unknown parameters,
${\bf n}({\bf r},k)$ denotes the residual noise, and ${\bf p}({\bf r},k;{\bf r}_0)$ is the planet signature.

\modif{The model can be built by placing two PSFs, one positive and one negative, properly positioned in the field of view to correspond to the field rotation at times $t_1$ and $t_2$ of a certain couple $k$. We assume here that the exposure time is short enough  that the companion does not undergo any azimuthal smearing, but it could be added if the smearing becomes significant.}
\modif{As we have only one representative PSF for all the images of the cube, we also assume that the PSF is known and does not vary with  time. As was true for other ADI approaches, if the instrument PSF does vary during the sequence, it  directly affects the companion flux estimate. The main limitation here is related to monitoring the PSF Strehl ratio along the sequence. If it were known, this knowledge could be included to rescale the searched companion signature in ANDROMEDA within the succesive differential images.
In the current paper, we also work under the hypothesis that the companion PSF core is temporally stable. In this case, the scaling factor $\gamma_k$ used to build up the differential images, which is related to the variability of the star halo, must be taken into account to build the model of the planet signature, }
\begin{eqnarray}
\mmodif{{\bf p}({\bf r},k;{\bf r}_0) = h({\bf r}-{\bf r'}(t_1,{\bf r}_0)) - \gamma_k \; h({\bf r}-{\bf r'}(t_2,{\bf r}_0))}
\label{eq1ter}
, \end{eqnarray}
where $h$ is the PSF of the system that can be estimated by providing an empirical PSF to the algorithm simply obtained by imaging the star (unsaturated image or without coronagraph). This PSF is shifted at a position $\mmodif{{\bf r'}(t_x,{\bf r}_0)}$ that the planet will have at a time $t_x$ if its position in the first frame is ${\bf r}_0$. \modif{If, however, it turns out that the intensity of the planetary signal and the PSF vary the same way in every image, the scaling factor should not be taken into account to build this model.} 

\modif{If another kind of method is used to perform the speckle subtraction, as suggested in Sect.~\ref{othersDI}, the model must be modified according to the subtraction performed. For instance, if the previously mentioned methods LOCI or PCA are used, the coefficients found to perform the subtraction must be recorded for each zone in order to build a consistent model.}

Since ANDROMEDA models \modif{differential images} through Eq.~(\ref{eq1bis}), the filtering procedure must also be applied to the reference PSF used to build the planet signature. In this way, the model remains consistent with the data and ANDROMEDA tracks the proper filtered planet signature in the filtered data sequence.


\subsection{Estimation of position and flux via maximum likelihood}
\label{max-likelihood}
ANDROMEDA's goal is to estimate the position and the flux of a potential planet orbiting around the target star using a probabilistic approach. This estimation is based on a MLE under the hypothesis that the residual noise inside the \modif{differential images is a Gaussian and white}. Far from the star, this hypothesis is fulfilled if the star PSF does not vary during the exposure because there is only photon and detector noise in the images \citep{Mugnier04}.
Closer to the star, quasi-static speckles are predominant and they follow  modified Rician statistics (MR) \citep{Goodman85,Soummer2004,Fitzgerald2006}, which are neither Gaussian nor totally white.

However, ANDROMEDA deals with high-pass filtered image differences.
First, it appears that the filtering process of the raw images increases the Gaussianity of the residual distribution in the images since it removes the slowly evolving structures while preserving the high spatial frequencies that evolve quickly. As the remaining noise \modif{inside an annulus is with a good approximation independant}, according to the central limit theorem, its distribution should follow a normal distribution.
Second, the speckle noise is temporally correlated from one frame to another. By performing the difference between close images, only the variation between the  speckle fields  in the two frames will remain in the \modif{differential} image. That is to say the created \modif{differential image} will contain the time derivative of the speckle and the \modif{power spectral density (PSD)} of a derivative tends toward a constant so the differential imaging process ``whitens'' the noise.
\modif{This whitening is a well-known consequence of the ADI \citep{Marois2008}, which is true mostly because we consider the reference frame of the sky. This property is verified empirically on real NaCo data in Sect. \ref{Sec-histo}.}
To conclude, the noise in the \modif{differential} images created to perform the MLE is, to a good approximation, Gaussian and white, thanks to the filtering and differential imaging performed. At the end of this section there is  further discussion of this hypothesis.

Under the assumption that the residual noise in the \modif{differential images} is white (in time and space), Gaussian, and non-stationary, the likelihood can be written
\begin{eqnarray}
 L(\mathbf{r_o}, a) \propto \exp \{ - {1 \over 2} \sum_k \sum_{\mathbf{r}} \frac{| \Delta(\mathbf{r},k) - a \, p(\mathbf{r},k;\mathbf{r_o})  | ^2 }{\sigma_\Delta^2 (\mathbf{r})} \}
.\end{eqnarray}
\modif{For each annular zone, at a certain distance from the star, this likelihood is computed for any possible initial position ${\bf r}_0$: the sum is made over all the $N_k$ couples found for this distance from the star and over all the pixels in the image field.}

ANDROMEDA analyzes the sequence of \modif{differential images} $\Delta({\bf r},k)$ by seeking the optimal values $(\hat{{\bf r}}_0,\hat{a})$
that maximize the logarithm of the likelihood,
\begin{eqnarray}
J({\bf r}_0,a) \triangleq \ln{L({\bf r}_0,a)} \propto -\sum_{{\bf r}, k} \frac{\left | \Delta({\bf r},k) - a \cdot {p}({\bf r},k;{\bf r}_0) \right |^ 2}{2 \sigma^2_\Delta({\mathbf{r})}}
\label{eq2}
.\end{eqnarray}
Equation~(\ref{eq2}) shows that maximizing the log-likelihood is equivalent to minimizing the sum of squared residuals between the \modif{differential image} and the model, weighted by the variance of the residual noise. We note  that the definition of the variance $\sigma^2_\Delta$ at Eq.~\ref{eq-variance} means that the weight is lower when there is a planet in the \modif{studied differential image} (and the noise is more overestimated to a higher degree when the planet is brighter and at closer separation). \modif{This bias induces an overestimation of the error, but this does not affect the estimation itself.}

The optimal flux value $\hat a$ for each possible initial position of the planet is easily computable analytically and is given by
\begin{eqnarray}
\hat{a}({\bf r}_0) =  \frac{\sum_{{\bf r}, k} {p}({\bf r},k;{\bf r}_0) \Delta({\bf r},k) /   \sigma^2_\Delta({\bf r})}{\sum_{{\bf r}, k}  {p^2}({\bf r},k;{\bf r}_0) / \sigma^2_\Delta({\bf r})}\;
\label{eqFluxest}
.\end{eqnarray}
This  equation shows that the estimated flux can be regarded as a cross-correlation between the planet signature and the \modif{differential image}, weighted by the noise variance averaged on every \modif{differential image}, the denominator being a normalization constant.
In \modif{the} ANDROMEDA code, after a processing per annuli, the data are combined in \modif{a single 2D map called a} flux map, giving at each pixel \emph{\emph{the flux of the object, if the object has this pixel position}}.

By inserting this expression of $\hat{a}({\bf r}_0)$ into the metric $J$, it is possible to obtain an expression for the log-likelihood that depends from now on only on the initial planet position ${\bf r}_0$,
\begin{eqnarray}
J^{\prime}({\bf r}_0) \triangleq J[{\bf r}_0,\hat{a}({\bf r}_0)]
= \frac{\left ( \sum_{{\bf r}, k} {p}({\bf r},k;{\bf r}_0) \Delta({\bf r},k) /   \sigma^2_\Delta({\bf r}) \right )^ 2}{\sum_{{\bf r}, k}  {p^2}({\bf r},k;{\bf r}_0) / \sigma^2_\Delta({\bf r})}\; + C
\label{eqJprime}
, \end{eqnarray}
where $C$ is a constant. The criterion $J^{\prime}({\bf r}_0)$ is a measure of the probability that there is  a point source at position ${\bf r}_0$ on the first image of the sequence. This formula is easily interpretable by saying that the planet has a high probability of being found at the position where the correlation between the model and the \modif{differential image} is the closest to one.
In practice $J^{\prime}({\bf r}_0)$ is computed for each possible initial position of the planet on a grid chosen as the original pixel grid of the images. The results are then a so-called likelihood map which \emph{\emph{has higher values at positions where the presence of a companion is more probable}}.

Once the likelihood and flux maps are computed, one purpose is to link the intensity of a likelihood peak to a probability of false alarm in order to assess whether it is indeed a true planetary signal. One way to proceed is to compute the standard deviation of the estimated flux, $\sigma[\hat{a}({\bf r}_0)]$, for each possible initial position, which describes how the noise propagates from the \modif{differential image} toward the flux map. The standard deviation of the estimated flux is given by
\begin{eqnarray}
\sigma^2[\hat{a}({\bf r}_0)] =  \left [{\sum_{{\bf r}, k}  {p^2}({\bf r},k;{\bf r}_0) / \sigma^2_\Delta({\bf r})}\right]^{-1}
\label{eqSigmaflux}
.\end{eqnarray}
It is then possible to define the signal-to-noise ratio of the estimated flux, \emph{\emph{S/N}}, as
\begin{eqnarray}
\mathrm{S/N}({\bf r}_0) \triangleq \frac{\hat{a}({\bf r}_0)} {\sigma[\hat{a}({\bf r}_0)]}\;
\label{eqSNRflux}
.\end{eqnarray}
The so-called \emph{detection limit}, that is to say the faintest detectable companion flux as a function of the separation from the star, at chosen threshold $N\sigma$ is thus given by $N \times \sigma[\hat{a}({\bf r}_0)]$. In other words, the standard deviation of the flux map is the detection limit at $1\sigma$.

It is straightforward to show that the \modif{S/N} can be rewritten as
\begin{eqnarray}
\mathrm{S/N}({\bf r}_0) = \frac{  \sum_{{\bf r}, k} {p}({\bf r},k;{\bf r}_0) \Delta({\bf r},k) /   \sigma^2_\Delta({\bf r})}{\sqrt{\sum_{{\bf r}, k}  {p^2}({\bf r},k;{\bf r}_0) / \sigma^2_\Delta({\bf r})}}
\label{eqJprimeSNR}
.\end{eqnarray}
\modif{Showing that the log-likelihood is simply the square of the S/N.} This means that maximizing the likelihood map is equivalent to maximizing the S/N map to find the most-likely position of a potential companion. The S/N thus gives a physical meaning to the log-likelihood $J^{\prime}({\bf r}_0)$.
Moreover, thresholding the S/N to zero is equivalent to computing $\hat{a}(\mathbf{r_0})$ under the constraint that it is non-negative, and the log-likelihood then obtained through Eq~\ref{eqJprimeSNR} incorporates this positivity constraint \citep[see][]{Mugnier09}. In summary, the likelihood and S/N maps contain the same information, and we chose to use the S/N map to perform detection since it directly yields the statistical significance of each detection.

\modif{It is then} possible to define a threshold $\tau$, to be applied to the S/N map, \modif{which corresponds} to a certain confidence level of the detection. Assuming that there is no planet \modif{anywhere} in the images (hypothesis $H_0$) our model shows that the \modif{differential} images equal the residual noise. If our hypothesis were strictly fullfilled, the residuals in the S/N map would have the same statistics as the noise in the \modif{differential} images (\modif{because each pixel of} the S/N map is a linear combination \modif{of pixels from the differential images} (via \modif{Eq.}~(\ref{eqFluxest}) and (\ref{eqSNRflux})), which is assumed to be white and Gaussian, with a probability density function (PDF), $p_{\text{S/N}}$ that is normal, zero-mean, and of unit variance. Under the assumptions on the noise, the \emph{\emph{probability of false alarm}} ($\text{PFA}$), defined as the probability that the S/N shows a signal above the chosen threshold $\tau$ under the hypothesis $H_0$, writes
\begin{equation}
\begin{split}
\text{PFA}(\tau) & = \mathit{Pr}(\text{S/N} > \tau )|_{H_o} \\
& =  1- \frac{1}{\sqrt{2 \pi}} \int_{-\infty}^\tau \exp(-\tau'^2/2) \; d\tau' \\
& = 1 - erf(\tau),
\end{split}
\label{PFA}
\end{equation}
where \emph{erf} is the so-called error function. The applicability and relevance of such a threshold $\tau$ is important as further discussed in Sect.~\ref{post-processing} and when applying ANDROMEDA over real data  in Sect.~\ref{sect-NaCo}.

However, these considerations are for the ideal case of a white, Gaussian, and non-stationary noise in the \modif{differential images}, for which the connection between PFA and threshold is perfect.
For real data, even if the filtering permits an increase in the Gaussianity of the noise and allows the differential imaging to remove an important fraction of its correlated component, the actual noise distribution still slightly deviates from our assumptions. The real noise distribution is closer to a MR distribution, which increases the PFA. Because of its structure, the MR distribution evolves radially: the quasi-static speckles have the same mean and variance only for a specific annulus. Close to the star, the speckle noise is dominant (and the noise is more correlated), whereas far from the star the photon noise, the readout noise, and the dark current prevail. The threshold should thus be increased according to the separation from the star where the companion is sought \citep{Marois2008}.

Another effect that is worth pointing out here is that for annuli close to the star, the field has not rotated as much as at large separations. At small separations, not only are there  a \modif{small} number of pixels contained in the annulus,   there are also \modif{few images} to be subtracted. These two effects reduce the number of degrees of freedom (in ANDROMEDA, the number of points inside the annulus, $\mathbf{r}$, plus the number of ADI couples, $k$) and corrupt the statistics: we are in a \emph{\emph{small sample statistics}} regime. In that case, \cite{Mawet2014} have shown, using a frequentist approach, that the PFA is  underestimated to an even greater extent. To correct for this bias, close to the star, the noise statistics  no longer follow a MR distribution, but a robust Student's t-distribution, which is valid only if the sample is independent and identically distributed  (which is the case if processes to whiten the noise, such as differential imaging, have been applied to the images). \modif{In practice, when correcting for this bias, we calculate the equivalent threshold that should be applied to provide the same PFA that  Gaussian statistics would give at the actual chosen threshold.}

As a consequence, the PDF chosen in Eq.~(\ref{PFA}) is no longer the normal law, $\mathcal{N}$, and should be replaced by the appropriate law: a MR PDF at large separation \citep{Marois2008} and a Student t-distribution PDF at small separation \citep{Mawet2014}. The detection limit calculation should then be modulated to take into account these two considerations. The next section is about how to take  the deviation into account from our model, in order to set a constant threshold throughout the whole S/N map. Setting a constant threshold enables the
probable point sources in the field to be automatically and systematically detected with a relevant confidence level.

\modif{We
note that the discrepancy between the model of noise made with respect to the real noise distribution does not affect the flux estimation in itself, but will affect the error on the flux estimation.}

\subsection{Post-processing: normalization of the S/N map}
\label{post-processing}
To complete the presentation of the method within this section, we can already indicate that the first tests performed on real data showed that the deviation from the noise model prevents one from applying a constant threshold to the S/N map and building an automatic procedure to detect companions. We can vizualize this deviation effect in Fig.~\ref{fig-SNRnorm}-\textit{left}, which notably shows a radial dependancy so the threshold should indeed be modified as a function of the separation.

This is not unexpected since if the noise model (white, Gaussian, and non-stationary) were consistent with the real values, the output S/N map from ANDROMEDA would have a zero mean and a standard deviation of one when, of course, excluding the pixels containing the signal from a companion. In this case, the threshold could be uniform all over the field. However, as explained in the previous section, in real images, the noise clearly deviates from this model and the standard deviation of the residual noise in the S/N map generally has a standard deviation larger than one \modif{which}  increases from the center to the edge \modif{and a mean of zero very far from the star that increases when going closer to the star}.

Knowing that we would like to obtain a S/N map that has a zero mean and a standard deviation of one, we propose a simple solution \modif{to preserving the PFA, which} consists in normalizing radially the S/N map by its own empirical radial standard deviation. \modif{This operation removes the radial dependency and allows one to apply a constant threshold to the S/N map, regardless of the distance from the star.} To normalize the S/N map, it is necessary to build an empirical radial profile of the standard deviation from the S/N map, without taking into account any peaks due to planetary signals. A method from \cite{Hoaglin1983} and \cite{Beers1990} to estimate this standard deviation, hereafter called \emph{\emph{robust standard deviation}}, has been implemented in ANDROMEDA. This method consists in calculating the median absolute deviation divided by a normalization factor enabling a robust estimate of the standard deviation. The factor is chosen so that the regular standard deviation and the robust value are identical in the case of a Gaussian distribution. Because only the global trend is needed, the mean radial robust standard deviation profile is smoothed over a certain number of pixels, $N_{smooth}$, which is a user-defined parameter that has to be set empirically according to the pixel scale and the size of the images.
Once a satisfactory profile that does not take into account the small variations from one annulus to its neighbor is obtained, the normalized S/N map is obtained by dividing the S/N map by this radial trend.

This method is reproducible and independent of the data set.
The post-processed S/N map thus gives the confidence level for a point source detection on each pixel of the image grid. Thresholding this map \modif{by a constant value throughout the field of view} provides a list of positions where a probable companion candidate is detected. As a result, in the following we  work exclusively with the normalized S/N map and from now on ``S/N map'' stands for the normalized S/N map. Through Eq. \ref{eqSNRflux}, as the S/N map has been normalized, the standard deviation flux map must also be normalized in a coherent manner to obtain a consistent relation between the normalized S/N and the standard deviation of the flux. \modif{The normalization process does not  significantly affect the position and flux estimations (the impact is lower than the given errors on these estimations) but normalizing too much (smoothing  the profile too much) artificially lowers the error on the flux estimation.}

\medskip
ANDROMEDA provides four 2D maps for analysis: the likelihood map, the S/N map, the flux map, and the flux standard deviation map. The two most useful outputs for detection and characterization are the S/N map, in which we can look for planetary signals and estimate their most likely position $\hat{r_0}$, and the flux map, in which we can read the corresponding estimated flux $\hat{a}(\hat{r_0})$. The map of the flux standard deviation is the detection limit at $1\sigma$. The following section deals with the analysis of the ANDROMEDA output to accurately compute the position and flux of potential companions present in the image field based on applications on real VLT/NaCo images to show concrete illustrations.


\section{Extracting companion information from the ANDROMEDA outputs}
\label{sect-Toolbox}
In this section we explain how to systematically and efficiently extract a list of likely companions with their subpixel position and estimated flux in terms of contrast. Once the S/N and flux maps are at hand, a complementary module is developed that automatically detects and analyzes the point sources present in the images. This module simply finds the signals above threshold in the S/N map and, knowing the expected shape of a planetary signal in the S/N map (Sect.~\ref{pattern}), assesses their subpixel positions and then estimates their corresponding flux thanks to the flux map (Sect.~\ref{positionToolbox}). Also, during the analyses, some tests are performed to discriminate probable planetary signals from artifacts (Sect.~\ref{artefactToolbox}).


\subsection{Resulting pattern from a planetary signal}
\label{pattern}
\begin{figure}[!h]
\centering
\resizebox{\hsize}{!}{\includegraphics{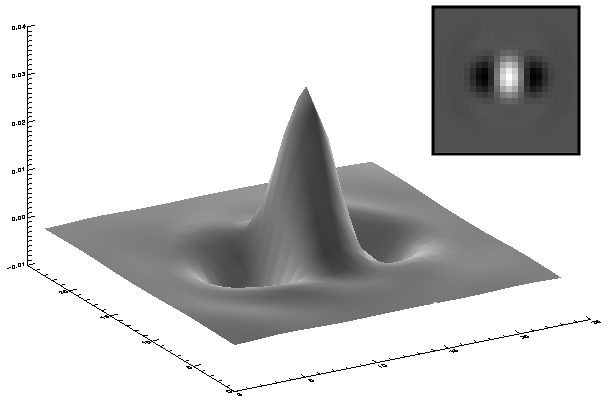}
                      \includegraphics{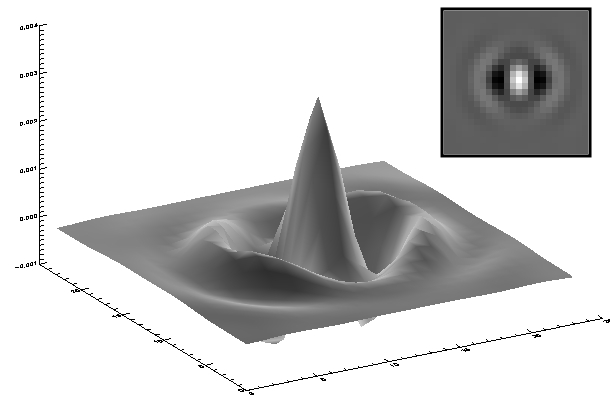}}
\caption{Illustration of the pattern (autocorrelation of the planet signature, Fig.~\ref{fig-PlanetSignature}) appearing in the maps as  evidence of a planetary signal.
\textit{Left:} Pattern obtained without filtering the raw images.
\textit{Right:} Pattern obtained with high-pass filtering of the raw images \modif{($\mmodif{F=1/4}$)}.}
\label{fig-pattern}
\end{figure}

If there is indeed a planetary signal, a specific pattern appears in both the S/N and flux maps. This pattern is well known and derived from Eq.~(\ref{eqJprime}) and Eq.~(\ref{eqJprimeSNR}): under the hypothesis that there is no noise in the data, it can be seen essentially as the autocorrelation of the planet signature $p(\mathbf{r},k;\mathbf{r_o})$ shown in Fig.~\ref{fig-PlanetSignature}, multiplied by the intensity of the candidate companion. Such a pattern is illustrated in Fig.~\ref{fig-pattern}, either with \modif{or} without pre-filtering of the data.

As expected, this pattern is made up of a main positive lobe surrounded by two negative lobes positioned perpendicular to the star-companion direction. The main lobe of the pattern is oval with its longest side along the radial direction with a typical length of $\lambda/D$, whereas the length of its perpendicular direction is dependent upon the chosen distance $\delta_{min}$.
In the case without pre-filtering, positively thresholding the S/N map retains only the main lobe. On the other hand, the filtering induces the appearance of two tertiary sickle-shaped positive lobes surrounding the main lobe as well as two tiny fourth negative lobes. Consequently, when applying a positive threshold, the tertiary lobes may remain in the thresholded map if the S/N is high enough.

\subsection{Identification, position, and flux retrieval of candidate companions}
\label{positionToolbox}
The S/N map permits the identification of the most likely companions and the extraction of their subpixel position in the first image.
The signals above the user-defined detection threshold are identified and classified according to decreasing peak S/N values. For each of them, a subwindow is extracted that fully encloses the central part of the candidate signal, as shown for instance in Fig.~\ref{fig-imagettePLUSmap}, and whose size is about $4\lambda/D$. To evaluate the subpixel position $\hat{r_0}$ of a candidate companion in a simple way, the planetary signal is interpolated to find the subpixel position of the maximum S/N, coinciding with the position of the planetary signal. The main lobe of the pattern expected from a planet is approximated by a Gaussian (this approximation is quite reliable in the very central part and less valid at the edges), and the subpixel position is identified at the maximum position of the 2D Gaussian fit of this main lobe. \modif{The error on the position estimation is simply given by the Gaussian fit uncertainty taken at $3 \sigma$. If needed, this error is corrected for the surrounding noise when it is too intense and might bias the error estimation.}

To read the corresponding estimated flux $\hat{a}(\hat{r_0})$, the same approach is used by performing a 2D Gaussian fit of the planetary signal, but this time using the flux map provided by ANDROMEDA. The extracted flux of the companion is then the value of the fitted Gaussian at the subpixel position, $\hat{r_0}$, previously retrieved on the S/N map. At this stage, ANDROMEDA returns the flux of the detection with respect to the flux of the input PSF. From this value, it is easy to derive the contrast between the two components by using all the information about how the input PSF has been imaged, such as the integration time of the PSF with respect to the integration time of each image of the cube, any normalization factor, or the neutral density transmission if one was used to image the star. We emphasize here  that this approach intrinsically takes  all the pre-processing applied to the data into account: the companion signature that is sought takes into account the pre-filtering, \modif{and then} the subtraction applied on image pairs separated by a small rotation difference. As a consequence the estimated flux does not depend on such parameters, and contrary to other approaches such as PCA or LOCI, \modif{there is no companion flux loss}. \modif{The error at $1\sigma$ on the flux estimation is directly read on the flux standard deviation map at the location $\hat{r_0}$ retrieved earlier. The error bars provided by the algorithm are the corresponding values at $3\sigma$.}


\subsection{Rejection of known sources of artifacts}
\label{artefactToolbox}
The method described so far, based on thresholding the S/N map, does not prevent the detection of artifacts whose signals are nevertheless above the threshold.
\modif{Any signal temporally varying in the images is not fully subtracted and may leave some high-level residuals in the differential images, which could mimic the expected planet signature. This kind of artifact appears in particular if the pupil field is not well stabilized or if the star is not well centered in the image during the observation.}
Therefore, it is needed to discriminate a posteriori  probable planetary signals from \modif{remaining} artifact signals. The rejection criterion setup is based on the morphological properties of the pattern expected from a planetary signal (Sect.~\ref{pattern}).

\begin{figure*}[t]
\centering
\resizebox{\hsize}{!}{\includegraphics{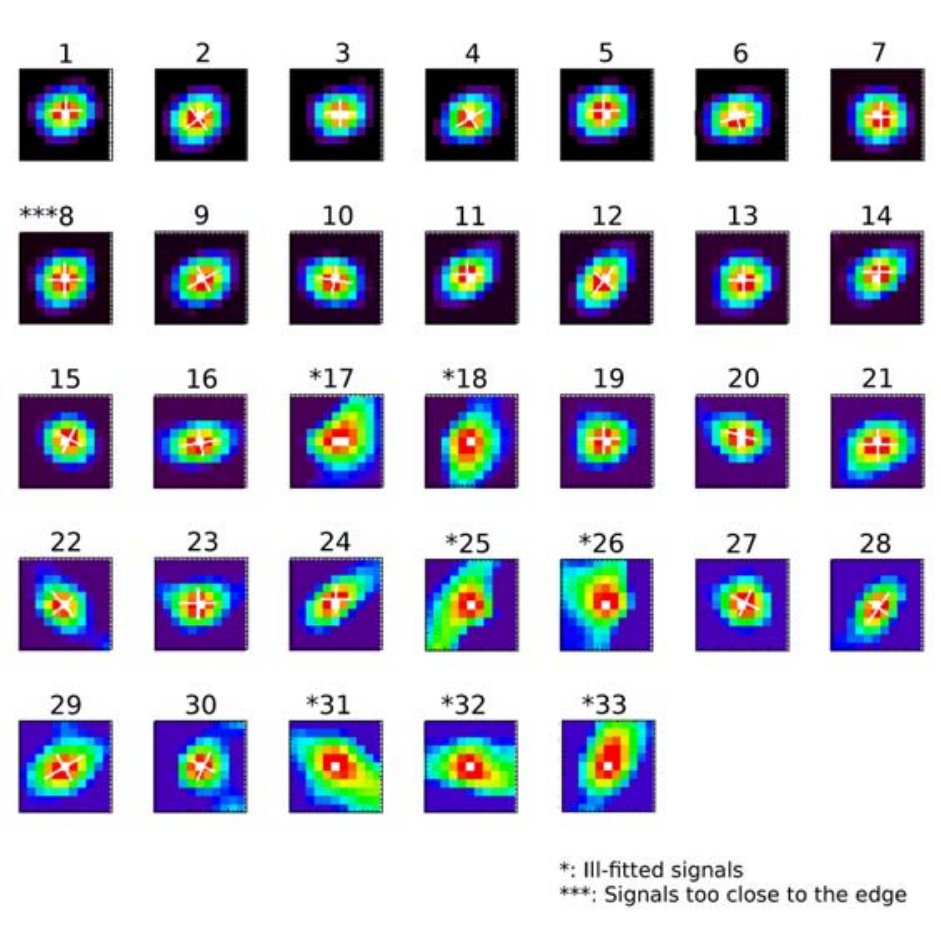}
                      \includegraphics[scale=0.75]{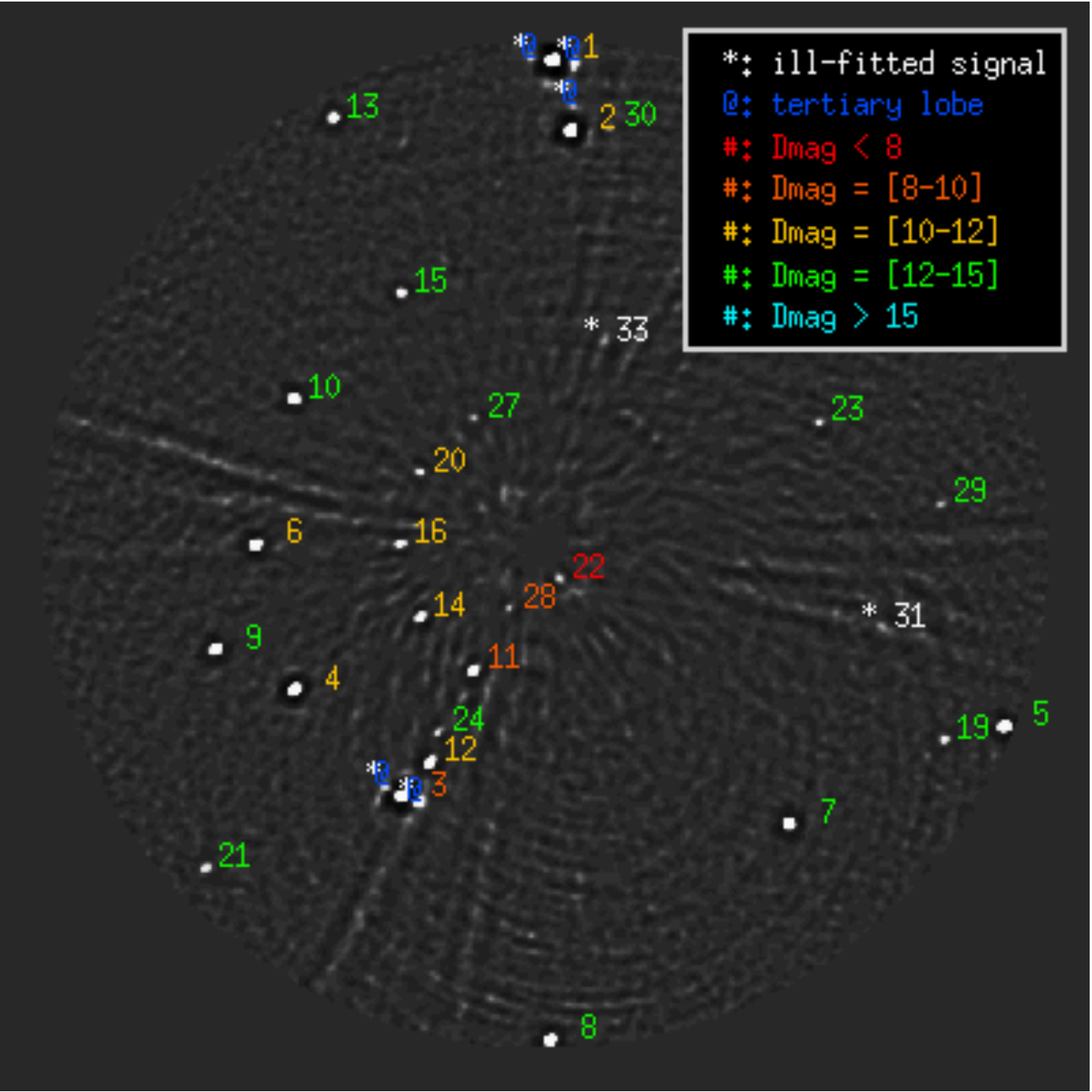}}
\caption{Images obtained by running the automatic detection module on the output provided by ANDROMEDA applied to the images of TYC-8979-1683-1 taken with the VLT/NaCo instrument (see Sect.~\ref{sect-NaCo}). The threshold is set at $5\sigma$.
\textit{Left:} S/N map subwindows ($11 \times 11$ pixels) in which the 2D Gaussian fits are performed for each detection. The subimages are classified by decreasing S/N and indexed by an increasing integer from one to the total number of detections above threshold. \modif{If the fit could not converge, one asterisk is placed above the corresponding subimage}. Otherwise, the two axes of the fitted Gaussian are superimposed and their lengths are equal to the FWHM.
\textit{Right:} S/N map showing the location of each detection in the first image (labeled by their index of detection) as well as their flux range (color of the index) and whether it is an artifact (one asterisk  means that the 2D Gaussian fit did not converge;  the @ symbol means that it is assessed as a tertiary lobe artifact).}
\label{fig-imagettePLUSmap}
\end{figure*}

Three main sources of errors have been identified that can be used to efficiently reject these false detections. To illustrate  these sources, practical examples are pointed out in the images of Fig.~\ref{fig-imagettePLUSmap} that were produced thanks to data provided by NaCo (see Sect.~\ref{sect-NaCo} for more details on the data processed). The subimages cut in the S/N map and classified by decreasing S/N signals are displayed in Fig.~\ref{fig-imagettePLUSmap}-\textit{left}.
Below are listed the three sources of false detection that have been diagnosed:
\begin{enumerate}
 \item The expected pattern contains tertiary positive lobes induced by the filtering procedure (Fig.~\ref{fig-pattern}-\textit{right}). If the S/N of the signal is high enough, it  these tertiary lobes may be above the threshold too and may thus be regarded as detections. These detected tertiary lobes are indeed surrounding only the very high S/N signals, such as  $\#1$, $\#2$,  and  $\#3$  visible in Fig.~\ref{fig-imagettePLUSmap}-\textit{right} via the dark blue symbol @. Three criteria are used to spot and reject these artifacts: their proximity to a high S/N signal, the S/N ratio between the regarded signal and its neighbor, and  their peculiar sickle shape which prevents the 2D Gaussian fit from converging. Such signals can be seen in Fig.~\ref{fig-imagettePLUSmap}-\textit{left} such as $\mmodif{\#18}$ (surrounding signal $\#1$) and $\mmodif{\#25}$ (surrounding signal $\#3$ - we note that some pixels are missing in the subimage because they had already been erased after detecting the contingent part of this tertiary lobe).
 \item The filtering may reveal the spider diffraction pattern in the S/N map that are not completely subtracted during the ADI (probably owing to the bad centering of the star or pupil tracking). Consequently, some signals can be found above the threshold that are actually artifacts due to \modif{this} diffraction pattern. However, if this is the case, the signal is quite elongated along the radial direction and it is thus possible to constrain the fit parameters to reject such signals. This kind of signal can be seen in Fig.~\ref{fig-imagettePLUSmap}-\textit{left}, such as signal $\mmodif{\#31}$, which is clearly located in the spider diffraction pattern visible in Fig.~\ref{fig-imagettePLUSmap}-\textit{right}.
 \item In areas where the speckle noise is quite high (close to the star), we might detect signals that originate from the speckle noise. This kind of signal is usually quite irregular compared to the expected pattern and can thus be rejected by constraining the 2D Gaussian fit to be neither too wide nor too shifted in the subimage. 
\end{enumerate}

Some other sources of instrumental error (hot pixels, etc.) can induce the detection of an actual artifact.
To find all these artifacts and reject them, three of the fit parameters are constrained (by using \modif{the IDL} function \texttt{mpfit2Dpeak.pro} \modif{developped by Craig B. Markwardt}, for instance): (i) The distance between the fit maximum and the center of the subwindow must be of less than a pixel, (ii) the FWHM of the fit must be consistent with the one expected from the pattern, and (iii) the long-axis orientation must be close to the radial direction. In the following, a detection whose fit has not converged or that does not respect one of the constraints listed above is referred as an ``ill-fitted'' companion and is therefore rejected.
In Fig.~\ref{fig-imagettePLUSmap}, when the 2D Gaussian fit is unsuccessful one asterisk is placed above the  subimage (such as the detections indexed $\mmodif{\#17}$ and $\mmodif{\#25}$). When the S/N decreases, the shape of the signals is less and less regular and is consequently often not well fitted. Indeed, the lower the S/N, the less probable it is for the signal to be real, hence the frequency of strangely shaped signals increases (e.g., we can see this effect in signal $\mmodif{\#31}$, which has a S/N value of $\mmodif{5.4}\sigma$ and lower). If the signal found is too close to the edge of the image, so its signal might be incomplete, it is impossible to obtain a correct 2D fit and in that case three asterisks are marked above the  subimage (e.g., signal $\#8$).

\section{Application to NACO data: Test case with TYC-8979-1683-1 data}
\label{sect-NaCo}

In order to assess the performance of the entire method, we ran ANDROMEDA on real data from the VLT/NaCo instrument. We applied ANDROMEDA to data collected by the NaCo Large Programme, which aimed at dectecting planets on wide orbits \citep{Chauvin2015}. The data we used consisted in observing a bright star surrounded by numerous background stars that are natural point sources acting as companions to be detected by the algorithm. Synthetic companions were then added inside the images in order to better quantify the ability of ANDROMEDA to accurately retrieve the position and contrast of the point sources present in the images.


\subsection{Data used for the test}
\label{data}
The data collected are sequences of saturated exposures (there is no coronagraph in that setup) taken in pupil tracking mode. The chosen star is TYC-8979-1683-1 \modif{(also called CD-62-657)} observed in 2011 on May 5 within the ESO program 184.C-0567(D) (PI: J.-L. Beuzit). This star is a G7 star of \modif{17} Myr ($V =9.36$, $H = 7.47$) located at \modif{75.6} pc from the Sun. The observation was made in H-band (filter centered around $1.66\mu m$) and stored in a $1024 \times 1024$ pixel frame (S13 camera inside the CONICA imager having a field of view of $13.6\arcsec \times 13.6\arcsec$)\modif{, the pixel scale being 13.22 mas/pixel}.

\subsubsection{Observation conditions}
\label{obs}
The star was observed during a total integration time of 36 min (giving 11 cubes of 29 frames each with an exposure time of 6.8 sec) and for a total field rotation of $18.5^\circ$. Seeing conditions were good but variable (seeing of 0.57\arcsec to 1.15\arcsec; coherence time of 4-9 ms; Strehl ratio of the reference PSFs: 21\% and 26\%).
\modif{The empirical PSF core FWHM is estimated to be of $\mmodif{4.75\pm0.05}$ pixels.}
\modif{The target was} observed close to meridian crossing, the PSF core is saturated inside a radius of 10-15 pixels (0.13\arcsec-0.20\arcsec), and integration times are set short enough so that the angular smear of potential companions is small, especially in the inner part of the field.
The data reduction of saturated exposures included sky subtraction (sky frame constructed from the median combination of exposures obtained at the \modif{five} different jitter positions: \modif{on minute timescales, the image center is moved by $\pm 3$ arcsec in x or y on the detector field.} ), flat fielding, bad pixels correction, and rejection of poor-quality exposures.
The location of the star center on each frame is determined by fitting the unsaturated portion of the saturated PSF with a 2D Moffat function. Individual frames are then shifted and registered to a common image center in between four pixels.

A short series of exposures with the  unsaturated star was taken before and after the main saturated sequence in order to build the reference PSF required as an input for ANDROMEDA. These unsaturated sequences were acquired with an exposure time of 1.7927 seconds in autojitter mode, using a neutral density filter (ND\_Short) of $1.19 \mmodif{\pm 0.05 \%}$ transmission factor. The first unsaturated image obtained for these data is shown in Fig.~\ref{fig-NacoData}-\textit{right}.

The parallactic angle associated with each frame of saturated images is computed from the observing time (converted from UTC to LST), assuming that individual exposures are recorded at constant time intervals within each data cube (time information is available only for the beginning and end of each data cube).

We thus obtained the three necessary inputs for ANDROMEDA: a reduced image cube, a PSF of reference, and a vector containing the parallactic angles of each image of the cube.

\subsubsection{Introduction of synthetic companions}
\label{synthetic-planets}
To better quantify and optimize the detection performance of ANDROMEDA using NaCo data, we implanted 20 additional synthetic substellar companions in the image cube. The signal of each synthetic companion was modeled using the unsaturated PSF image and inserted in the individual reduced frames, taking into account the field rotation that occurred between the exposures. The procedure used to inject synthetic planets in the images of the data cube is explained in \cite{Chauvin2012}.

\begin{figure}
\centering
\includegraphics[scale=0.55]{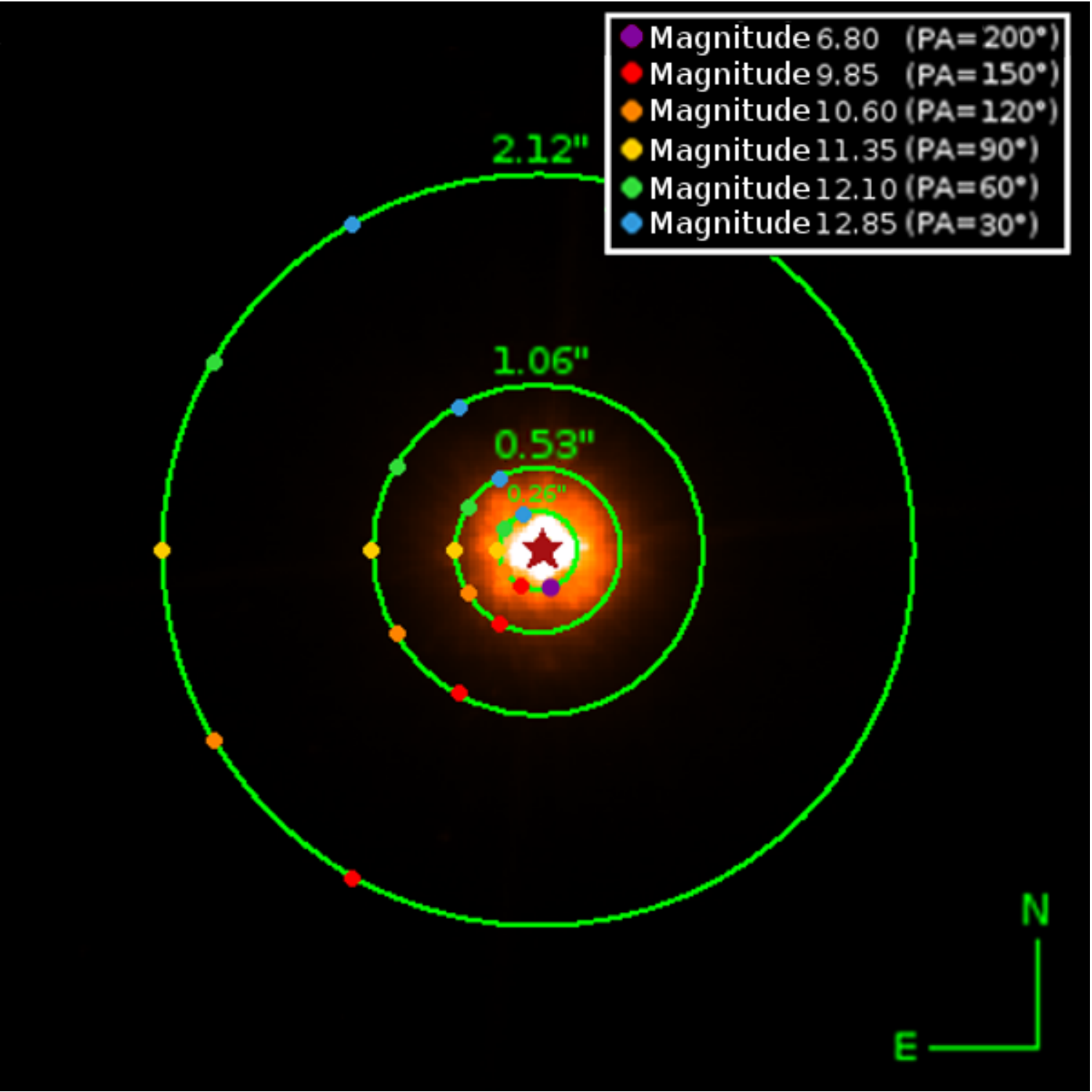}
\includegraphics[scale=0.14]{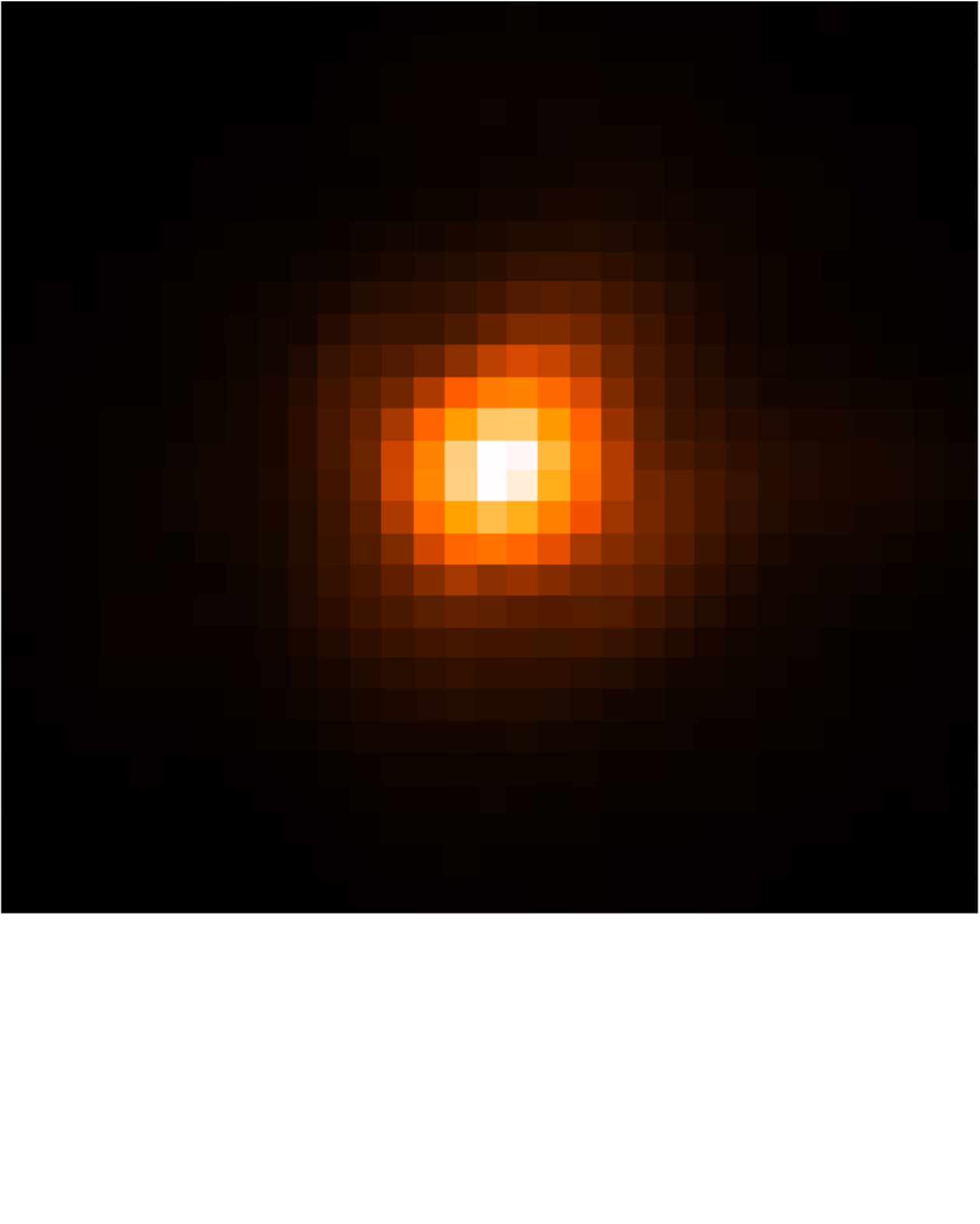}
\caption{NaCo data of TYC-8979-1683-1. \textit{\modif{Top:}} One reduced saturated image of the raw data cube ($\mmodif{600 \times 600}$ pixels, linear scale). The position of the injected synthetic planets is indicated with colored dots on the green circles.
\textit{\modif{Bottom:}} Reduced unsaturated image used as a reference PSF to perform ANDROMEDA ($32 \times 32$ pixels).}
\label{fig-NacoData}
\end{figure}

The twenty synthetic companions were introduced along five radial directions of respective position angles of $30^\circ$, $60^\circ$, $90^\circ$, $120^\circ$, and $150^\circ$ on the first image for each of these angle, at five angular separations of 0.26\arcsec, 0.53\arcsec, 1.06\arcsec, and 2.12\arcsec. The synthetic companions of the same position angle are of equal flux, each with peak intensities corresponding to magnitude differences of respectively \modif{12.85}, \modif{12.10}, \modif{11.35}, \modif{10.60}, and \modif{9.85} for the five position angles in increasing order. An example of one saturated image of TYC-8979-1683-1 on which the location of these introduced synthetic companions are added is shown in Fig.~\ref{fig-NacoData}-\textit{left}. \modif{In order to verify the detection capability at close separation, we also added a companion at 0.26\arcsec, with a PA of $220^\circ$ and a magnitude of 6.8 (contrast of about $2.10^{-3}$)}.

\subsection{ANDROMEDA's output}
We applied ANDROMEDA to the data described above and obtained the four output detailed in Sect.~\ref{sect-method}. To process these data, we have chosen the user-defined parameters values gathered in Tab.~\ref{tab-paraNaco} and whose suitability is discussed in detail in Sect.~\ref{sec-ParaSens}.
\begin{table}[!h]
\caption{User-defined parameter values chosen to run ANDROMEDA on the TYC-8979-1683-1 VLT/NaCo data.}
\label{tab-paraNaco}
\centering
\begin{tabular}{l c}
\hline \hline
\textbf{User-defined parameter} & \textbf{Used value} \\
\hline
Filtering fraction, $F$                              & 1/4 \\
Minimum distance for subtraction, $\delta_{\rm min}$ & $1.0 \; \lambda$/D \\
Width of the annuli for ADI, $dr$                    & $1.0 \; \lambda$/D \\
Ratio optimization to subtraction area, $R_A$        & 2 \\
Size of the input PSF window, $N_{psf}$              & $32 \times 32$ pixels  \\
Normalization profile smoothing, $N_{smooth}$        & $18$ pixels   \\
\hline
\end{tabular}
\end{table}

\begin{figure*}[t]
\centering
\resizebox{\hsize}{!}{\includegraphics{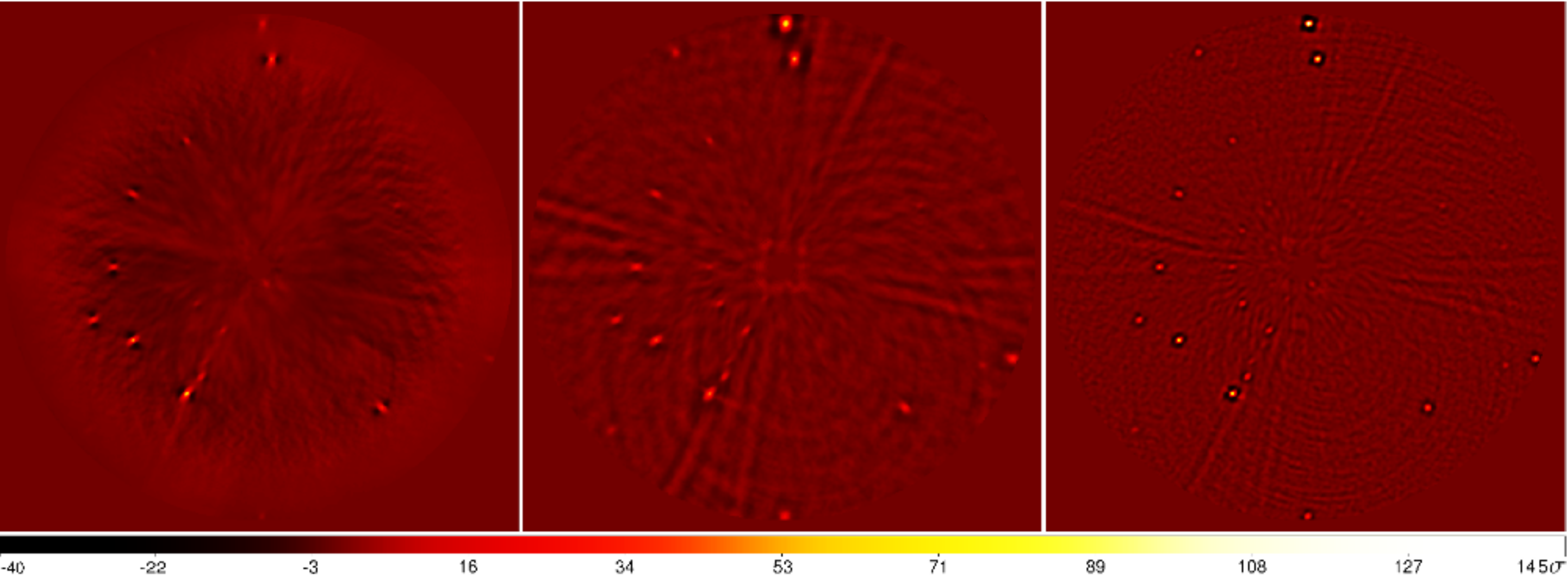}}
\caption{Effect of the pre-processing (spatial high-pass filtering of the raw data) on the output S/N map obtained by processing the TYC-8979-1683-1 NaCo data (including both real and injected synthetic companions) with ANDROMEDA. \textit{Left:} S/N map obtained without pre-filtering. \textit{Middle:} S/N map obtained with high-pass pre-filtering, using $F=1/16$. \textit{Right:} S/N map obtained with high-pass pre-filtering using $F=1/4$. \modif{All these S/N maps are normalized following the procedure in Sect.~\ref{post-processing}}}
\label{fig-SNRmapFilter}
\end{figure*}

The S/N map obtained is shown in Fig.~\ref{fig-SNRmapFilter} according to different filtering fractions. It can be seen that when the low frequencies are removed, the   companion candidates appear more distinctly, mostly those close to the star. The diffraction patterns induced by the spider of the telescope are revealed as two blurry lines crossing at the center (their cone shape is due to the ADI process per annulus and shows the expected pattern, negative-positive-negative); the patterns remain \modif{despite} the ADI procedure, probably due to the smearing of the star in the field.  If not filtered enough, the number of false detections increases, especially close to the star, and many companions that are close to the star and faint are missing. Moreover, the accuracy of the estimated position and flux decreases since the planetary patterns are wider and less regular. We consequently chose to set the filtering fraction to $F=0.25$: a quarter of the low frequencies are removed. This  value is a trade-off between not losing too much companion flux (see Sect.~\ref{pre-processing}), detecting as many true companions as possible, and decreasing the overall number of false alarms. We
note that  the oversampling of this data set is of $1.6$, and so the filtering fraction must be smaller than $F=0.6$.

\begin{figure*}[t]
\centering
\resizebox{\hsize}{!}{\includegraphics{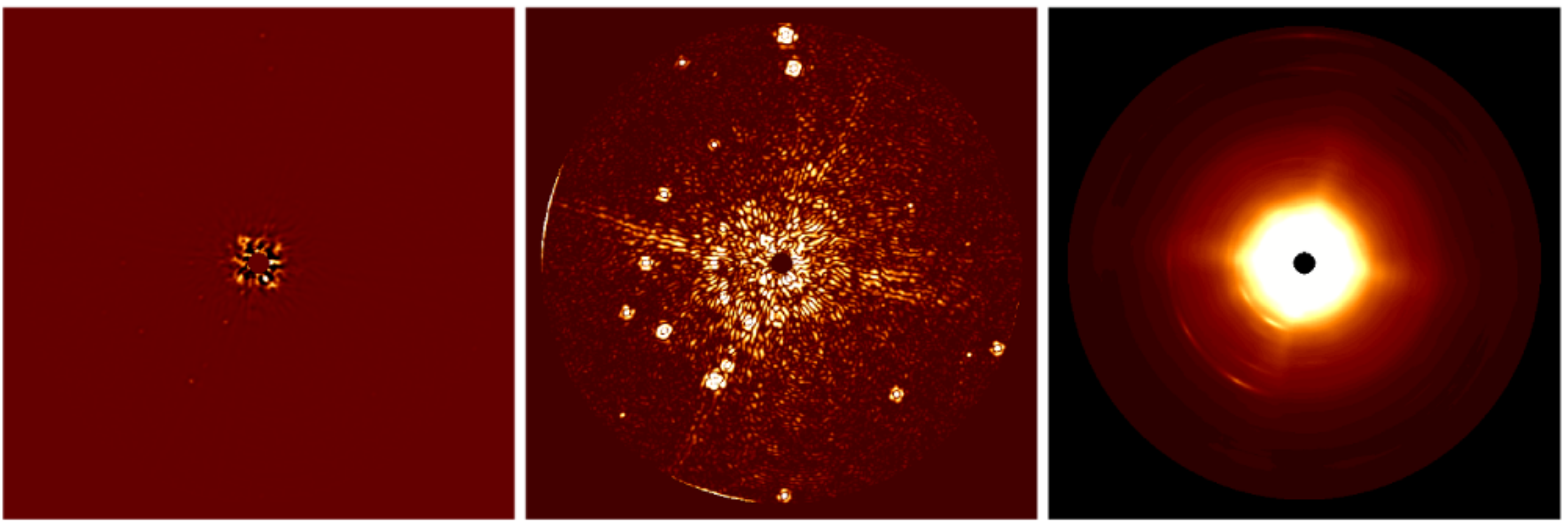}}
\caption{Output obtained by processing the TYC-8979-1683-1 NaCo data (including both real and injected synthetic companions) with ANDROMEDA.
\textit{Left:} Flux map. \textit{Middle:} Likelihood map. \textit{Right:} Map of the standard deviation of the flux. For each of these maps (as for the S/N maps in Fig.~\ref{fig-SNRmapFilter}) the central region corresponding to the star has been masked. The intensity scale is logarithmic for the likelihood and the standard deviation of the flux maps, and linear for the flux map. White corresponds to the highest value and black to the minimum value.}
\label{Fig-andromeda-output}
\end{figure*}

In Fig.~\ref{Fig-andromeda-output}, from left to right, are the flux map, the likelihood map, and the standard deviation of the flux map obtained with the parameters listed in Tab.~\ref{tab-paraNaco}. As expected, the flux and the likelihood maps look similar to the S/N map, and the standard deviation of the flux shows a clear radial decreasing trend from the center to the edge, as does the residual speckle noise distribution in the \modif{differential} images.


\subsubsection{Effect of the normalization process}
\begin{figure*}[t]
\centering
\resizebox{\hsize}{!}{\includegraphics{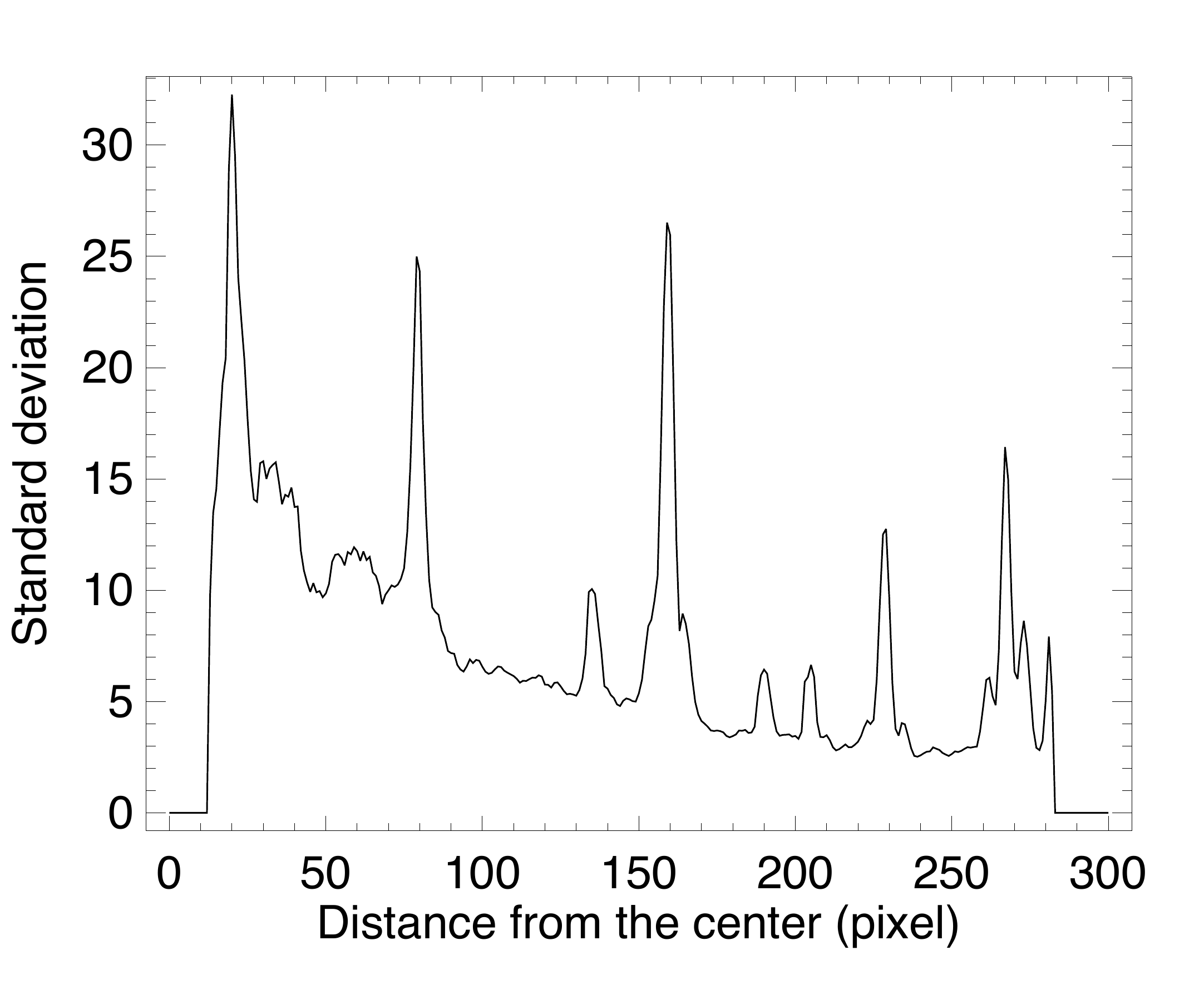}
                      \includegraphics{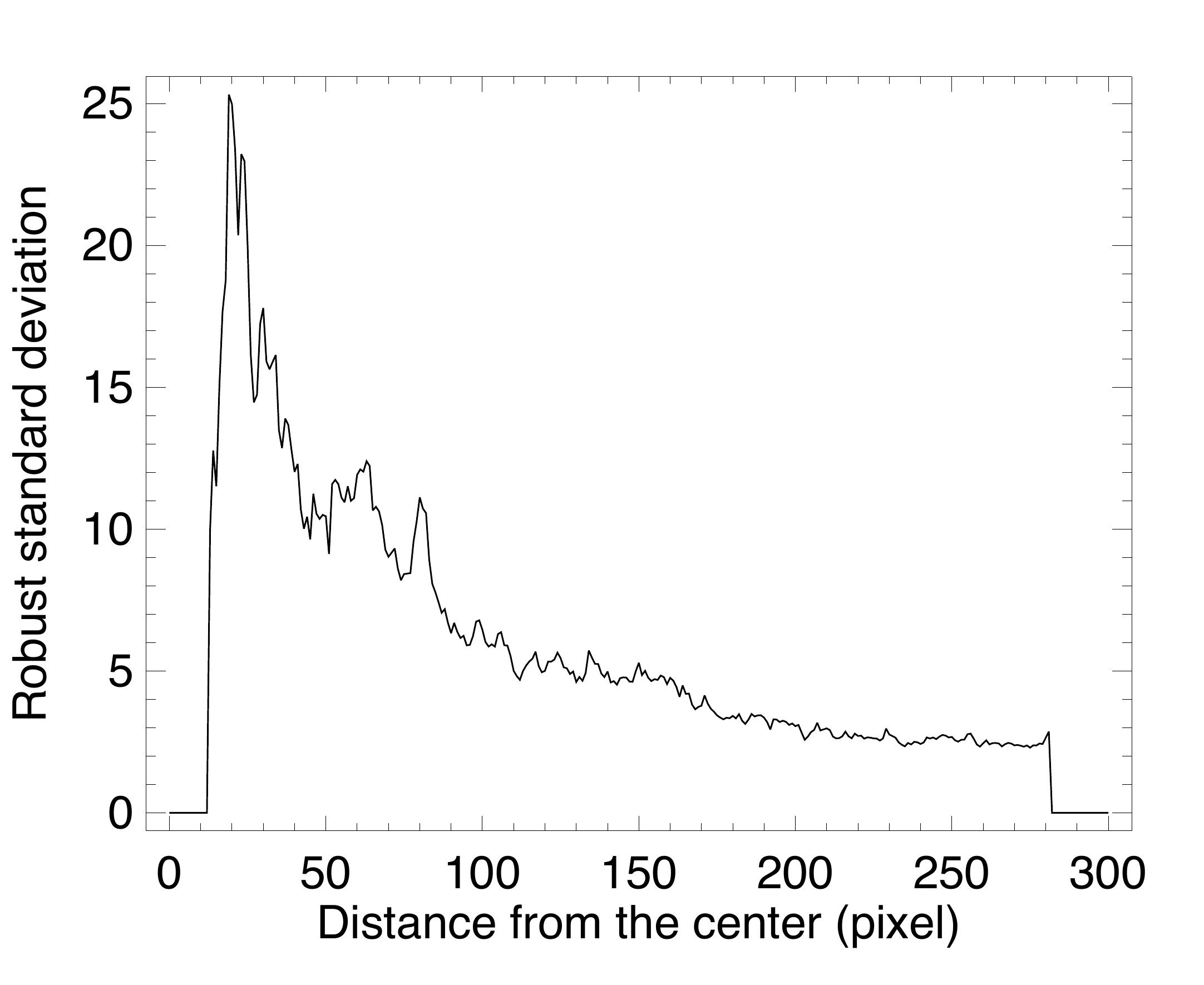}
                      \includegraphics{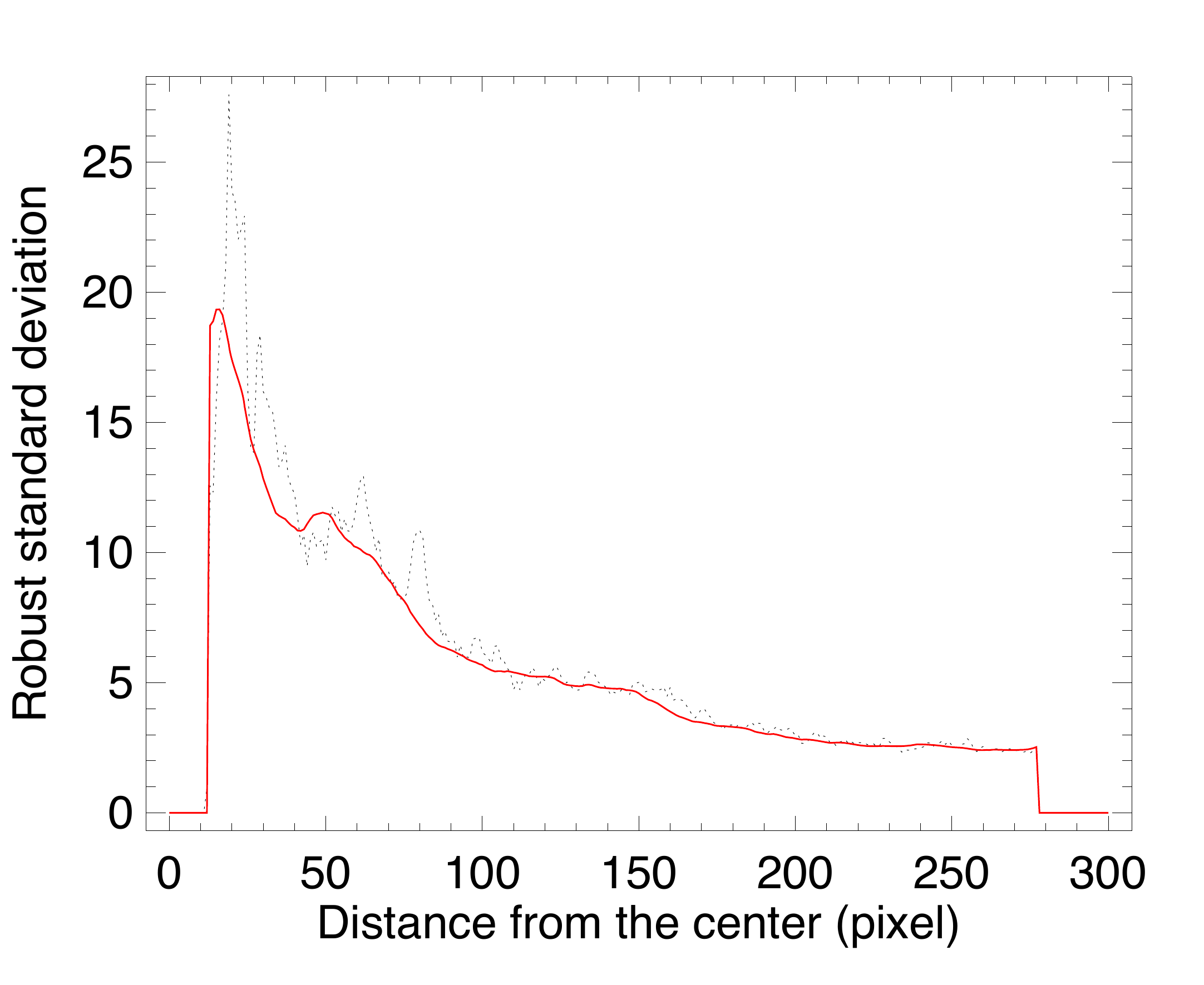}}
\caption{\modif{Azimuthal} mean standard deviation of the S/N map as a function of the distance to the star TYC-8979-1683-1.
\textit{Left:} Regular radial profile of the S/N standard deviation. We can visualize the high peaks indicating the presence of companions at certain separations.
\textit{Middle:} Radial profile of the S/N robust standard deviation. The highest peaks have disappeared but the profile is still jagged.
\textit{Right:} Radial profile of the S/N robust standard deviation smoothed over \modif{28} pixels to obtain the global trend of the S/N map standard deviation deprived of its companions.
\modif{The ANDROMEDA process reduces the exploitable zone between the IWA (at $13$ pixels here since $IWA = 4\lambda/D$) and the OWA (at $281$ pixels here since $OWA= \text{Size}_{image}/2 - N_{PSF}/2 - dr$).}}
\label{fig-SNRnormSteps}
\end{figure*}

In order to correctly normalize the S/N map, we have plotted the radial profile of the robust standard deviation per annulus of the S/N map, according to different smoothing values. We then chose the best profile, the one that  smoothed tiny irregularities and kept the global trend. Some profiles are presented in Fig.~\ref{fig-SNRnormSteps} to compare the robust versus regular standard deviation and smoothed versus unsmoothed profiles, thus justifying the choices that were made to build the normalized S/N map.

\begin{figure*}[t]
\centering
\resizebox{\hsize}{!}{\includegraphics{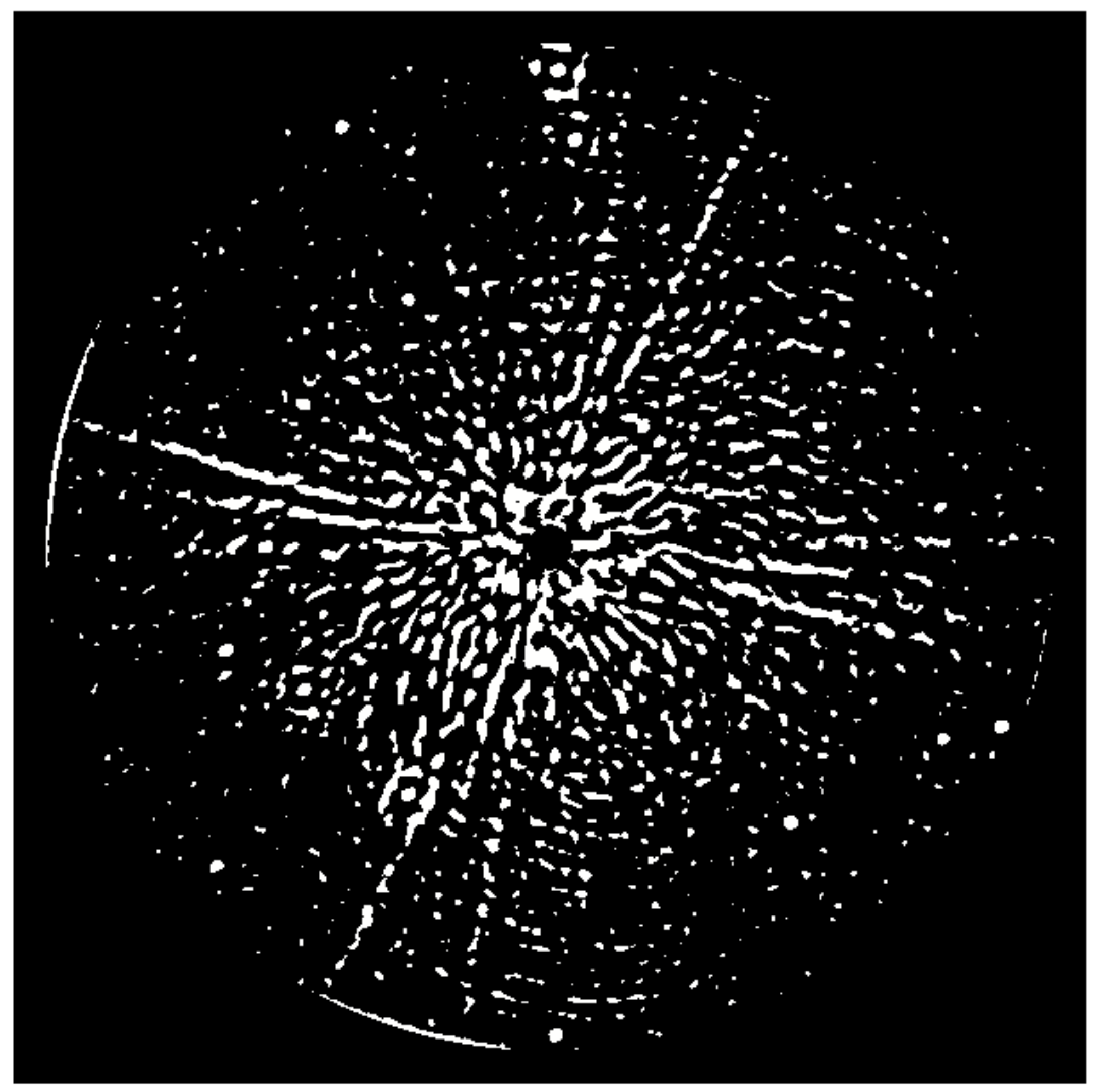}
                      \includegraphics{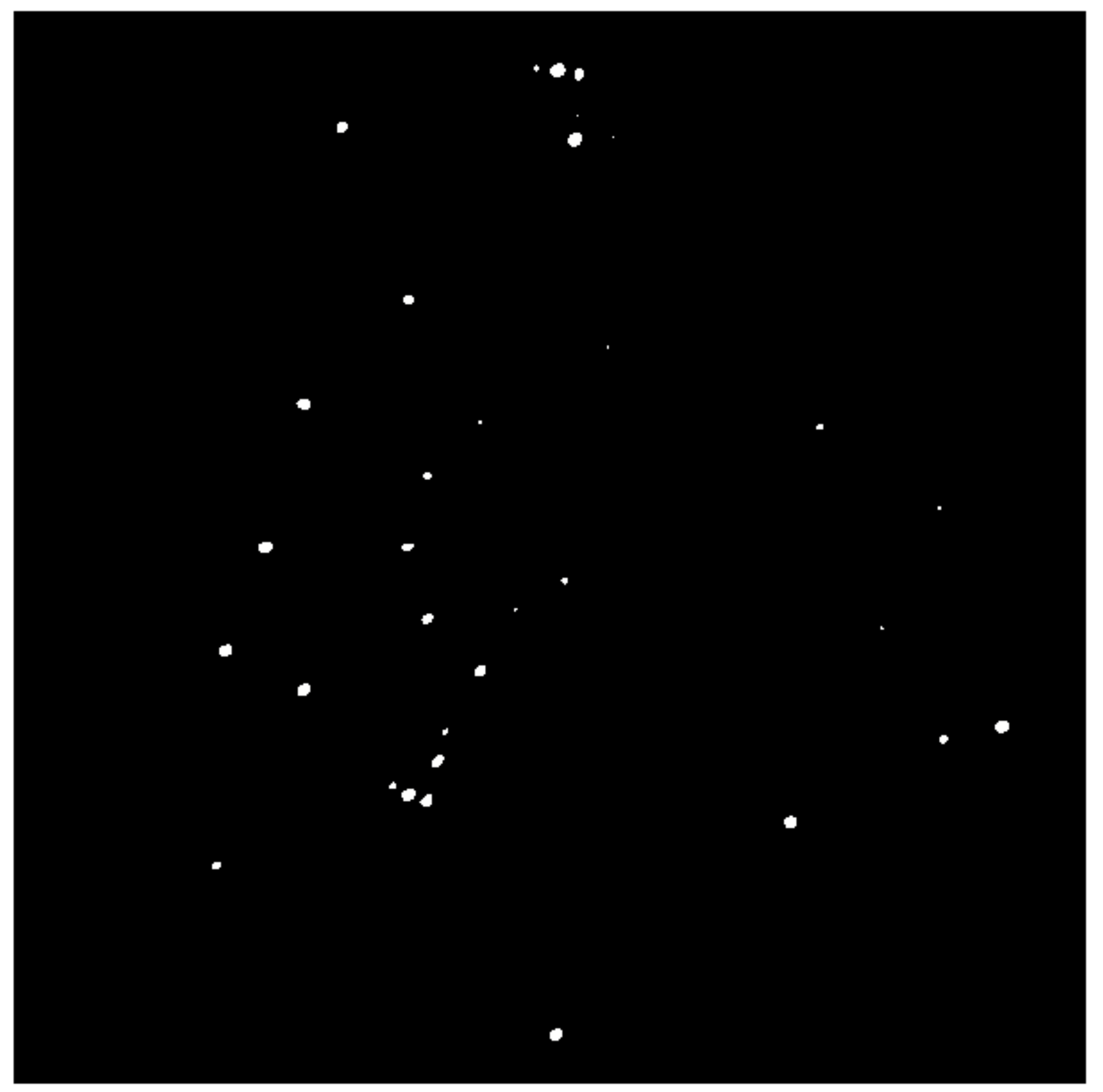}}
\caption{S/N maps that have been thresholded to 5$\sigma$.
\textit{Left:} Non-normalized thresholded S/N map. We observe that many signals are above threshold (white), mostly close to the center and along the spider diffraction pattern.
\textit{Right:} Normalized thresholded S/N map. Here only the probable planetary signals are found above the threshold. It is possible to perform an automatic detection on this map.}
\label{fig-SNRnorm}
\end{figure*}

The maps in Fig.~\ref{fig-SNRnorm} show the raw S/N map (left) and the normalized S/N map (right) both thresholded to $5\sigma$. The non-normalized S/N map (left) does not provide any workable information and just illustrates that its radial profile decreases toward the edge; however,  the normalized S/N map  reveals only the probable point sources in the map and therefore enables the automatic detection of companions.


\subsubsection{Interest of the filtering and normalization process}
\label{Sec-histo}
In order to verify the consistency of the filtering and normalization procedure, we plotted the histograms of the residuals in the S/N map. This permits us to check if the PDF of the residuals has the expected shape, that is to say, Gaussian and white. 

\begin{figure*}[t]
\centering
\resizebox{\hsize}{!}{\includegraphics{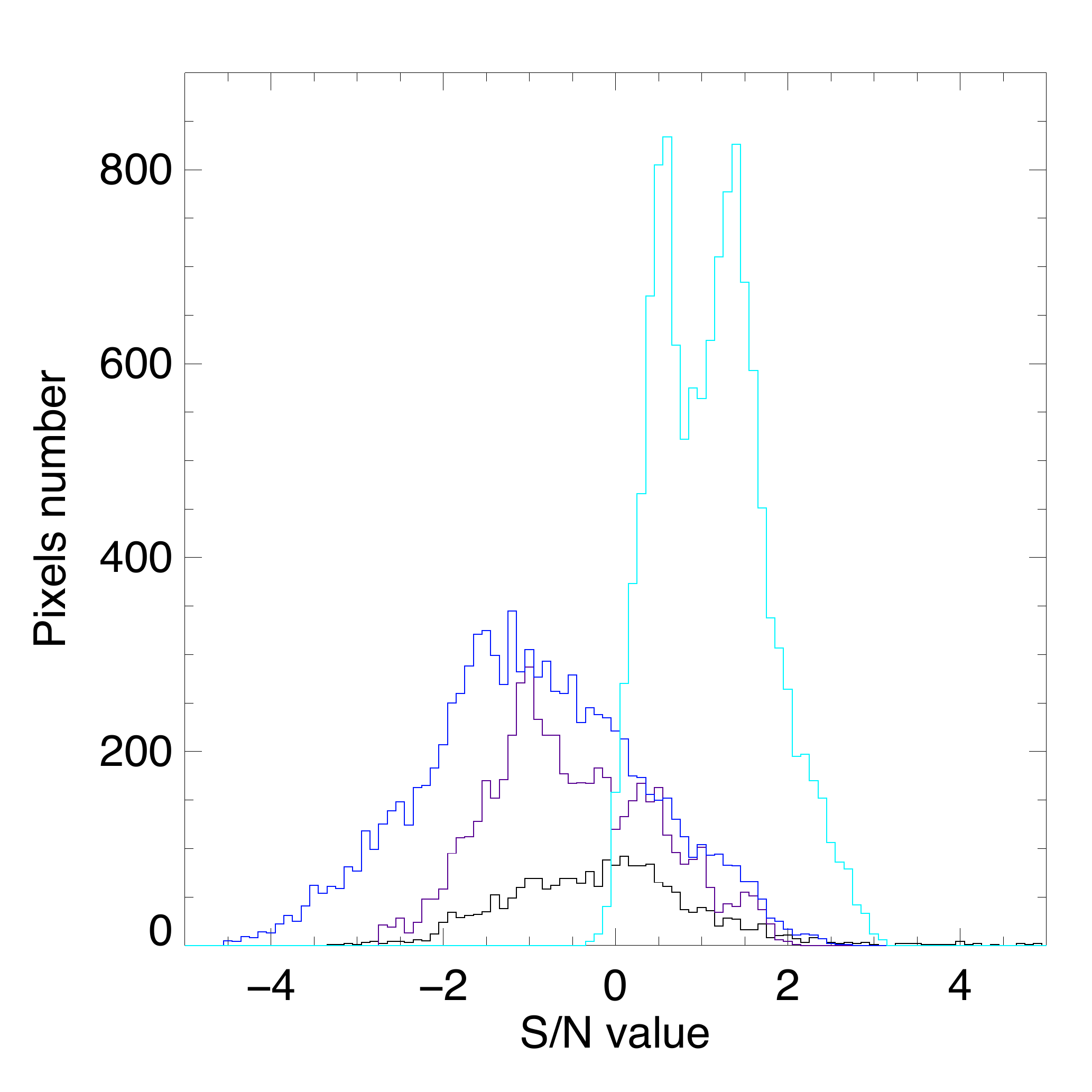}
                      \includegraphics{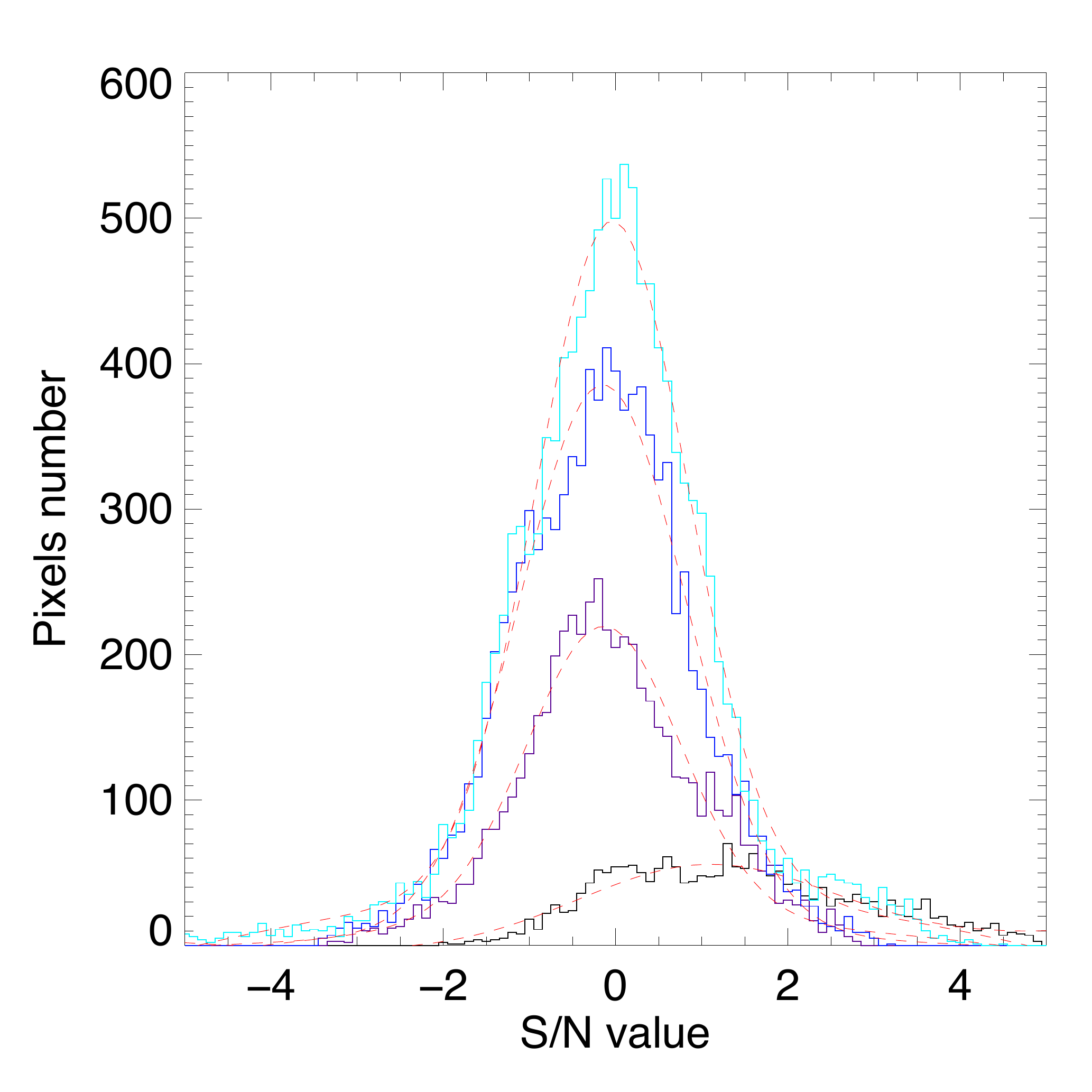}
                      \includegraphics{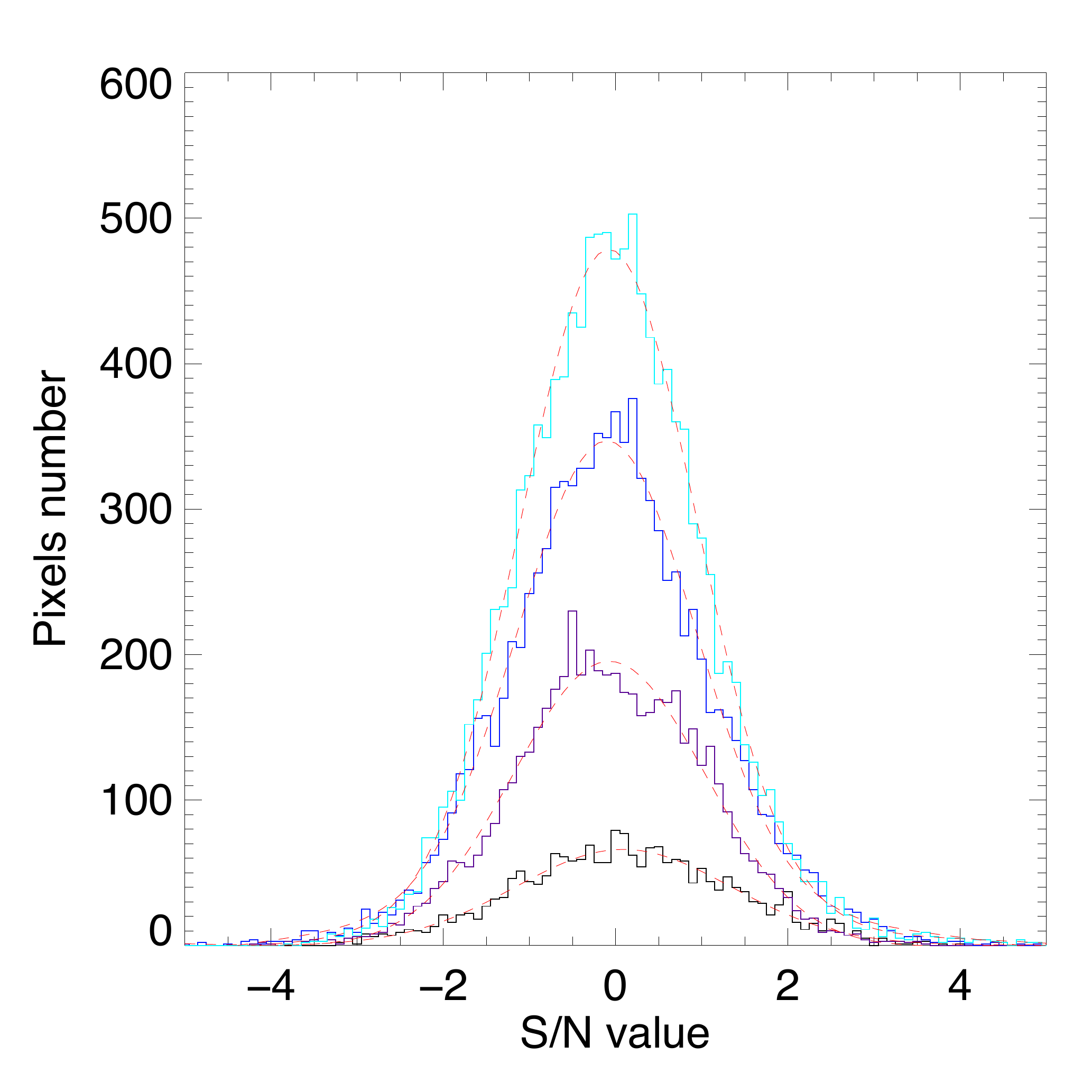}}
\caption{Histograms of the residuals in the normalized S/N map, obtained using the NaCo data of TYC-8979-1683-1, inside \modif{four different annuli centered on the star: black solid lines are for an inner radius of 15 pixels ($\sim \mmodif{5}\lambda/D$), purple solid lines for 50 pixels ($\sim \mmodif{17}\lambda/D$), dark blue solid lines for 95 pixels ($\sim \mmodif{32}\lambda/D$), and light blue solid lines for 247 pixels ($\sim \mmodif{90}\lambda/D$). Each annulus has a width of 15 pixels except for the largest, which  only has a width  of 8 pixels. No obvious planetary-like signal can be found inside these annuli.}
\textit{Left:} without filtering.
\textit{Middle:} with a pre-filtering, using $F=1/16$.
\textit{Right:} with a pre-filtering, using $F=1/4$.
\modif{Gaussian fits of the histograms are overplotted in red dashed lines}.}
\label{fig-histoFilter}
\end{figure*}

The histograms shown in Fig.~\ref{fig-histoFilter} have been produced from the normalized S/N map obtained by processing the filtered images with ANDROMEDA on TYC-8979-1683-1, using the parameters gathered in Tab.~\ref{tab-paraNaco}. It can be checked that   the residual noise in the S/N-map has been made Gaussian thanks to the high-pass filtering (Sect.~\ref{pre-processing}) and also that, thanks to the normalization process, this Gaussian distribution is centered on zero, which means that the residuals  have a mean value of zero, and its FWHM is equal to two, which corresponds  to a standard deviation of about one.

\subsubsection{Impact of the user-defined parameters}
\label{sec-ParaSens}
This part is a short discussion of the influence of each of the user-defined parameters on  ANDROMEDA's output. The optimal values of the user-defined parameters found from these investigations are set as default in the algorithm, though it is appropriate only for this particular TYC-8979-1683-1 data set. The parameters are discussed in the order they are used in the algorithm. We note here that most user-defined parameters do not have a major impact on the estimation performance. Their optimal values \emph{for this data set}, as well as the strength of their impact in terms of detection are gathered in Tab.~\ref{tab-parameters}.

\paragraph{Performing ADI:}
To perform the ADI according to the technique explained in Sect.~\ref{ADI}, the parameter $\delta_{min}$ must be chosen carefully to avoid self-subtraction of the companion while subtracting \modif{images as temporally correlated as possible}. \modif{For this data set, the} best compromise is obtained when using $\delta_{min}=1.0 \lambda/D$.

Since the \modif{displacement between two given images} varies with the separation from the star, the ADI is performed over an annulus of user-defined width $dr$ that should be the thinnest. Indeed, there are slightly fewer artifacts detected when $dr$ decreases. \modif{However, the width of the annuli should not be too small because it increases the processing time. A compromise is to set $dr=1\lambda/D$ as default.}

\modif{When choosing to perform the image subtraction with either a  least-squares fit or a L1-affine fit, the width of the annulus over which} $\gamma_k$ is calculated has a certain ratio $R_A$ with respect to the width $dr$ of the subtracted annuli (see Sect~\ref{ADI}).  There is a compromise to make when choosing the optimization area so that it can avoid discontinuities (being thicker) while minimizing the flux difference within the annulus taken into account (being close to the subtraction area). As expected, when $R_A$ is too large many artifacts may be detected, but when $R_A$ is too close to one there are discontinuities appearing between adjacent annuli. Under these considerations, the optimal value is set to $R_A=2$.

\paragraph{Maximum likelihood:}
The size of the PSF window, $N_{psf}$, must completely enclose the full signal, otherwise it induces strong errors on the flux estimation. On the other hand, the larger the window, the longer it takes to run ANDROMEDA. Thus, this parameter can be optimally set so that the first bright ring fits inside the PSF window.

We
note that the four parameters talked about previously ($\delta_{min}$, $dr$, $R_A$, and $N_{psf}$) are parameters that have been pointed out from the first developments of the ADI method. Some algorithms such as TLOCI \citep{Marois2014-TLOCI} or PCA \citep{Amara2012,Soummer2012}, which are improvements of ADI, try to reduce the number of parameters \modif{for this part of the algorithm}.

\paragraph{Normalization of the S/N:}
The number of pixels on which the S/N robust radial standard deviation profile is smoothed, $N_{smooth}$, is another important parameter.  Too much smoothing biases the normalization and provokes the appearance of artifacts detected at less than 0.5\arcsec\ from the star but that can be easily rejected \textit{\emph{a posteriori}}. On the other hand, a low smoothing may result in missing faint close signals. Thus, a good value is to smooth the profile over about $28$ pixels, but this value is strongly dependent upon the \modif{oversampling} and upon the quality of the images. It is the only parameter that must \modif{currently} be chosen by hand after visualizing the trend of the profile with different values. \modif{We note that the chosen value does not need to be very precise: for this data set, it can vary from about 26 to about 50 pixels without adding too many artifacts above threshold}.
\begin{table*}
\centering
\caption{User-defined parameters set as defaults in the ANDROMEDA pipeline and their respective significance.}
\label{tab-parameters}
\begin{tabular}{l l c c c }
\hline \hline
\textbf{Parameter} & \textbf{Definition} & \textbf{Units} & \textbf{Default value} & \textbf{Impact} \\
\hline 
$F$     & Filtering fraction  (Sect.~\ref{pre-processing})              & -                & 1/4  & low \\
$\delta_{\rm min}$ & Minimum separation to build \modif{the differential images}  (Sect.~\ref{ADI}) & $\lambda$/D     & 1.0  & high \\
$dr$    & Width of annuli on which ADI is performed  (Sect.~\ref{ADI})  & $\lambda$/D      & 1  & low \\
$R_A$   & Ratio optimization to subtraction areas  (Sect.~\ref{ADI})   & -                 & 2    & low \\
$N_{psf}$       & Size of the square PSF image (H-band filter of NaCo)  & pixels            & 32   & see text \\
$N_{smooth}$    & Smoothing of the S/N robust standard deviation profile  (Sect.~\ref{post-processing}) & pixels      & 18 & high  \\
\hline
\end{tabular}
\tablefoot{The right column shows the impact of the user-defined parameters over the whole process: If \emph{low}, the value can remain fixed and if \emph{high}, it should be tuned according to the data set.}
\end{table*}

To conclude, along with a correct normalization, ANDROMEDA provides workable output that allow  an automatic detection procedure to be built. The results obtained with the automatic procedure, developed to extract the position and flux of the candidate companions from the output maps, are presented in the next section.


\subsection{Analysis of the planetary signals detected}
\label{sec-NacoEst}
In a second step, the automatic detection module was applied on the maps, using a threshold of $5\sigma$.
The subimages extracted from the S/N map in which the fits are performed are shown in Fig.~\ref{fig-imagettePLUSmap}-\textit{left}. On account of the pixel scale of the imaging camera, the size of the subwindows is set to $11\;pixels$, which is the expected planetary pattern size ($\sim 4 \lambda/D$) that  encloses the whole planetary signal. The S/N map on which the position of the detected signals is visible and labeled by their index of detection is shown in Fig.~\ref{fig-imagettePLUSmap}-\textit{right}.

The program detected $\mmodif{39}$ signals above this threshold, including $\mmodif{25}$ reliable detections (Sect.~\ref{artefactToolbox}). Of course the number of detections found in the S/N map depends on the threshold set; this dependence is dealt with in detail in Sect.~\ref{sec-ThreshSens}. 
As expected, only the highest S/N signals are found surrounded by tertiary lobes that are detected (detections $\#1$, $\#2$, and $\#3$, having S/N values of respectively $\mmodif{150}\sigma$, $\mmodif{105}\sigma$, and $\mmodif{95}\sigma$). It is also noticeable that, as expected again, all the fits that have converged in the S/N map have also converged in the flux map and vice versa. Thanks to the criteria classiying the detections laid down in  Sect.~\ref{artefactToolbox} , the automatic detection procedure efficiently separates probable true companions from artifacts.


\subsubsection{Performance in terms of detection and analysis}
\label{performance}
To better quantify for the performance of the method in terms of detecting point sources present in the image field and in estimating their positions and flux, it is possible to use the knowledge we have about the injected synthetic planets. Knowing their exact position and contrast, it is possible to compare \modif{them} with the values obtained by the algorithm. It is also important to check that all the planetary companions found are indeed above the detection limit derived from \modif{this data set}.

\begin{figure*}[t]
\centering
\includegraphics[scale=0.45]{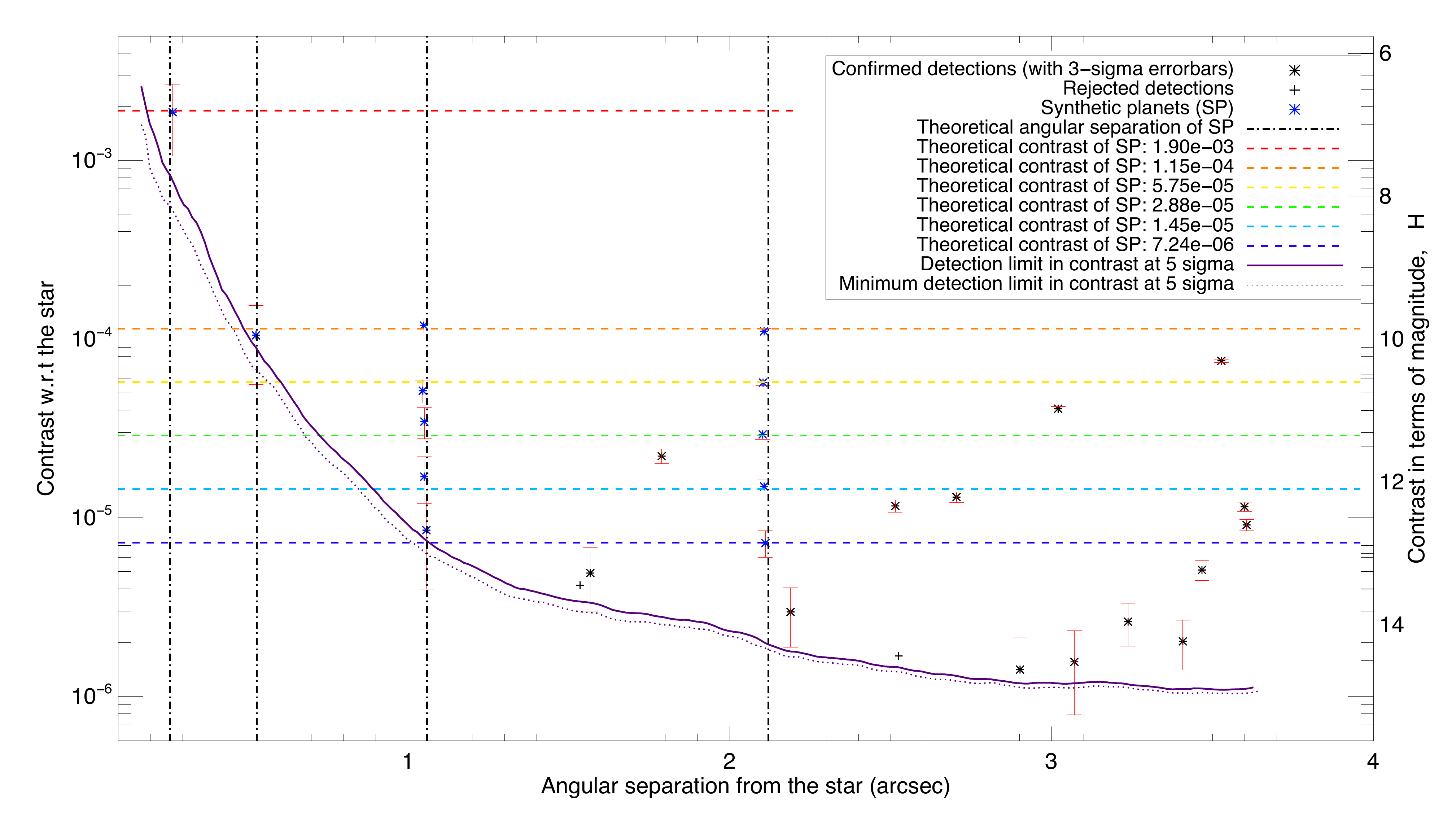}
\caption{Contrast of the detections as a function of their distance to the star TYC-8979-1683-1. The theoretical contrasts and distances of the 20 injected synthetic planets are shown as vertical and horizontal lines: we should have found a planetary signal at every crossing. \modif{One synthetic companion has also been added at the closest separation just above the detection limit in order to check its consistency (red dashed line).} The detection limit \modif{at $5\sigma$} is overplotted  (solid line). This detection limit curve is corrected for the underestimation of the small sample statistics  at close separation from the star (see Sect.~\ref{max-likelihood}). \modif{The minimum value of the flux standard deviation map at $5\sigma$ for each separation is overplotted for information (dotted line). The radial median of the flux standard deviation map at $5\sigma$, however, provides a realistic estimation of the detection limit}. The detected signals assessed to be tertiary lobes are not shown on the graph.}
\label{fig-perf}
\end{figure*}

Figure~\ref{fig-perf} shows the contrast in terms of magnitude of the detected point sources as a function of the angular separation between the detected signal and the central star. On this graph, \modif{six} horizontal lines represent the original contrast of the synthetic companions and four vertical lines are placed at their theoretical separation from the star. Consequently, we know that signals from the synthetic planets ought to be found at every crossing between the horizontal and vertical lines, except for the upper horizontal line, which only has  one synthetic companion at the closest separation. The detection limit is \modif{overplotted on the graph (solid line) and 
the error bars in terms of $3\sigma$ error on the contrast estimation are added to the graph.}

We first notice that, \modif{apart from the brightest synthetic companion that we injected just above the detection limit, none of the five synthetic planets that had been injected at 0.26\arcsec\ from the star is detected}. Moreover, only the brightest synthetic planet located at 0.53\arcsec is detected. This result is expected since each of these undetected synthetic planets is located under the detection limit curve. The detection limit shows the following trend: close to the star only bright companions can be detected, whereas going farther from the star enables  fainter signals to be detected. Of course, the detection limit is dependent upon the chosen threshold, but even with a very low threshold, it is impossible to detect these faint and close signals. As \modif{all} image processing methods, ANDROMEDA is limited by the observation conditions.

The estimated positions of the detected synthetic planets match the theoretical values, even very close to the star. The error bars in position due to the Gaussian fit uncertainty at $3\sigma$ are of about $2.0$ mas (from $0.02$ mas for the highest signal to $6.0$ mas for the faintest). These errors in position are not shown in Fig.~\ref{fig-perf} since \modif{they are} smaller than the symbol size.
As expected, the flux is better estimated when the companion is bright and far from the star. A good agreement is still observed between the theoretical contrast and the estimated value, \modif{knowing that the error bars shown in Fig.~\ref{fig-perf} are only the ones given by the map of the flux standard deviation but without taking into account the instrumental errors or the algorithm's intrinsic errors}.

To assess the errors intrinsic to the algorithm, one approach is to slightly move the user-defined parameters (which  might influence
the estimations) from their optimal value. The boundaries within which such user-defined parameter are made \modif{to vary} are the following: $\delta_{min} \in [0.2;2.0]\lambda/D$, $dr \in [0.5;3]\lambda/D$, $R_A \in [1;4]$, and $N_{smooth}$ must be varied experimentally from no smoothing to what can be reached with the data at hand. In this way, we obtained an error of about $5.0$ mas in position and of about $0.25$ magnitude\modif{, depending on the intensity of the signal}. \modif{Accounting for this deviation, the real position and contrast values of the injected fake companions are within the error bars of the ANDROMEDA estimations}.

In exoplanet imaging, \modif{the dominant errors on the position and flux estimations are usually due to instrumental sources}. But in the case of poor quality data, it must be verified which of the instrument calibration errors or the algorithm unstability errors are dominant.


\subsection{Threshold sensitivity}
\label{sec-ThreshSens}
\begin{figure}[!h]
\centering
\resizebox{\hsize}{!}{\includegraphics{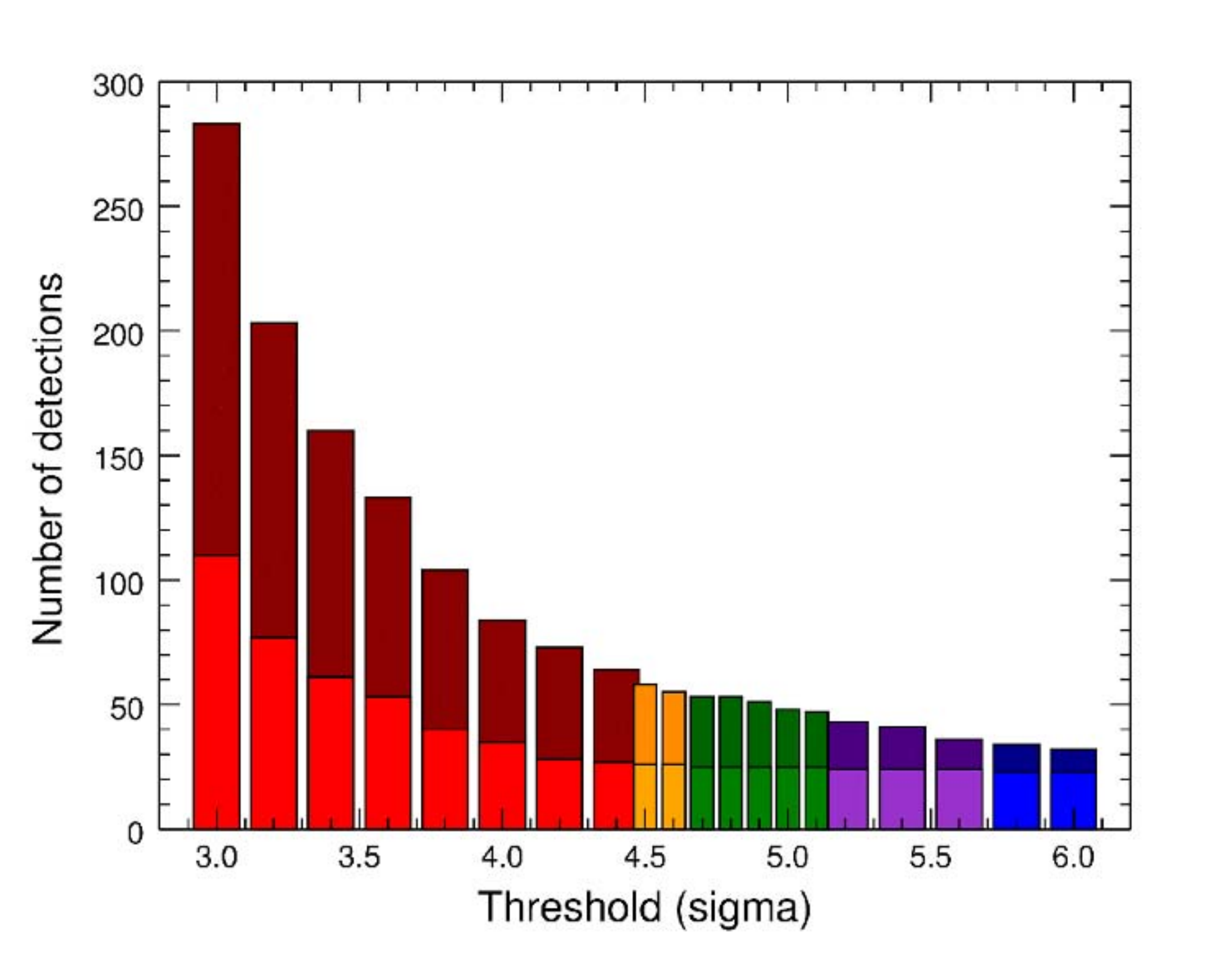}}
\caption{Number of total detections \modif{(dark color)} and of confirmed detections \modif{(light color)} as a function of the threshold set for the TYC-8979-1683-1 NaCo images. \modif{The optimal zone (where all  25 companions are detected) is shown in green. A false alarm zone is shown in red (more than one false detection) and orange (one false detection) and a loss zone in blue (more than one true companion is not detected) and purple (one true companion is not detected)}.}
\label{Fig-Thresh}
\end{figure}

This section  discusses the optimal threshold \modif{range} that would reveal \modif{as many } true companions as possible, while not missing any.
\modif{In particular, it is tested whether a constant detection threshold is efficient all over the field of view. This discussion relies upon a study of the behavior of the number of detections as a function of the threshold. Discussing missed detections and false positives requires assumptions on the actual number of companions present in the field. The series of fake companions is a firm basis. The number of astronomical background stars produces additional test cases to check the detection capability homogeneity, and can be tested in comparison to other ADI approaches. We assume in the following that the exact number of detectable point sources for this data set is $25$.} The graph in Fig.~\ref{Fig-Thresh} was obtained \modif{by running ANDROMEDA} under the same conditions as before (Tab.~\ref{tab-paraNaco}), and represents the number of detections as a function of the threshold which  varies from $\mmodif{3\sigma}$ to $6\sigma$. On this graph, both the total number of detections and the number of so-called reliable detections are indicated.

This graph shows the existence of a short optimal threshold range, between $4.7\sigma$ and $5.1\sigma$, for which the number of detections is exactly the \modif{one suspected of being the true one}. \modif{When increasing the threshold above this range, the faintest signals (like $\mmodif{\#39}$) and/or the ones very close to the star (like $\mmodif{\#33}$) are the first to be missed by the automatic detection process, whereas several tests have proved that these are most likely to be real point source signals.} When decreasing the threshold, some detections that are actually not true signals are detected and \modif{still} pass the test that is supposed to sort out artifacts from potential real signals. \modif{These detections are usually faint signals ($\Delta H\sim$ 14-16), located far from the star where the algorithm should show better performance, that are most likely to be false alarms due to residual speckles.} \modif{Moreover, a careful examination of the shape of this graph brings useful insights to help the user choose the threshold. An initial   inspection shows that lowering the threshold increases  the number of detections exponentially and that there is a plateau, from $\sim 4.5\sigma$ to $\sim 5.0\sigma$, where the number of detections is quite stable as it is for the number of true detections. Also, by looking at the ratio of true detections versus ill-fitted ones, below a threshold of $4.5\sigma$, the number of ill-fitted detections becomes significantly higher than the number of true detections, whereas above the threshold this ratio remains quite constant. This study is of course strictly valid only for this data set, which has the advantage of containing many point sources, but it should still give a fair idea of the behavior of the algorithm as a function of the threshold.}

\medskip
Even if this test case is very advantageous, the goal of this section is to check the overall method skills. This procedure proved to be very efficient in terms of detection ability and in terms of position and contrast estimations with low error bars. We noticed that thresholding the S/N map was not enough to provide only real probable planetary signals so we added an a posteriori classification  of the detections, which also proved its efficiency in discriminating artifacts from very probable true point source signals. We note  that ANDROMEDA does not provide more information than can be reached given the observation conditions (according to the detection limit trend), but the algorithm does provide, within some minutes, reliable information in terms of position and flux of the detected companions. Of course where the noise is higher (toward the star), the estimation is slightly less accurate, but the errors still remain under the observational errors such as the PSF centering or the tip-tilt correction.  Section~\ref{sect-betapic} is a comparison of the results obtained by processing the well-known case of Beta-Pictoris using the PCA methods or ANDROMEDA.


\section{Results on $\beta$ Pictoris and comparison with the PCA-KLIP method}
\label{sect-betapic}
In addition to testing the method on the previous field, populated with both numerous background stars and additional synthetic companions, we also applied it \modif{to} the emblematic and well-known case of $\beta$ Pictoris. This star \modif{is surrounded} by a debris \modif{disk} inside which only one planet, $\beta$ Pictoris $b$, has been detected by imaging. This close companion is located at \modif{$\sim 9$} AU from its host star and was first detected by \cite{Lagrange09}.
This companion has since  been observed repeatedly because of its outstanding astronomical interest for many reasons. It is a moderate-mass giant planet at close physical separation, making it a good candidate for having been formed within the disk. Its interactions with the remaining debris disk can still be witnessed and dynamical constraints can also be derived from the orbital motion and from numerous falling evoporating bodies as observed in spectroscopy \citep[see][]{Lagrange1988,Beust1990}. In the context of this paper, \modif{observations} of this companion \modif{provide} an excellent test case of our method in the challenging case of the detection of a companion at very short separation ($<0.5$\arcsec) with appropriate high-quality coronagraphic images; the images processed by ANDROMEDA should be compared to those processed thanks to other methods representing the \modif{state of the art}.

The $\beta$ Pictoris data on which we applied ANDROMEDA are described in \cite{Absil2013}. The goal of this data set was to \modif{search} for closer planetary companions (down to 2~AU from the star), which would be at the origin of the dynamical excitation in the planetary system. This search required the use of a newly developed coronagraph that allows detections at angular separation as close as \modif{$0\farcs1$}.

\subsection{Data used for comparison}
$\beta$ Pictoris \modif{was} observed on 31 January 2013 for 3.5h at the Very Large Telescope (Chile) using the NaCo instrument (ESO programme ID 60.A-9800). The images were recorded in the $L'$ band (centered on $3.8\mu$m), which is commonly used to work in a more favorable planet/star contrast regime \modif{compared to} shorter wavelengths, and to obtain an image quality close to that delivered by extreme adaptive optics (XAO) systems in the near-infared. To minimize the impact of \modif{starlight} in the images, the new\modif{ly commissioned} $L'$-band Annular Groove Phase Mask \citep[AGPM,][]{Mawet2005, Mawet2013} vector vortex coronagraph \modif{was} used. This kind of coronagraph provides \modif{an} inner working angle and \modif{an} achromaticity over the bandwidth of interest \modif{as good as any other} phase mask developed so far.

The data are constituted of ten blocks of 200 successive frames of 0.2s, each taken under fair turbulence conditions.
The observing sequence was obtained in pupil-tracking mode, showing a total field rotation of $83^\circ$. Two non-coronagraphic \modif{PSFs} were taken before and after the observing sequence by \modif{moving the star far from} the \modif{vortex center}. \modif{The pixel scale of the camera  used is $27.15$ mas/px and the PSF FWHM is empirically measured to be 4 pixels.} All the information concerning the data set used in this section and the pre-processing steps applied before running image processing algorithms such as ANDROMEDA (basic cosmetic treatment, recentering, frame selection, etc.) can be found in \cite{Absil2013}.

\begin{figure*}[t]
   \centering
   \resizebox{\hsize}{!}{\includegraphics{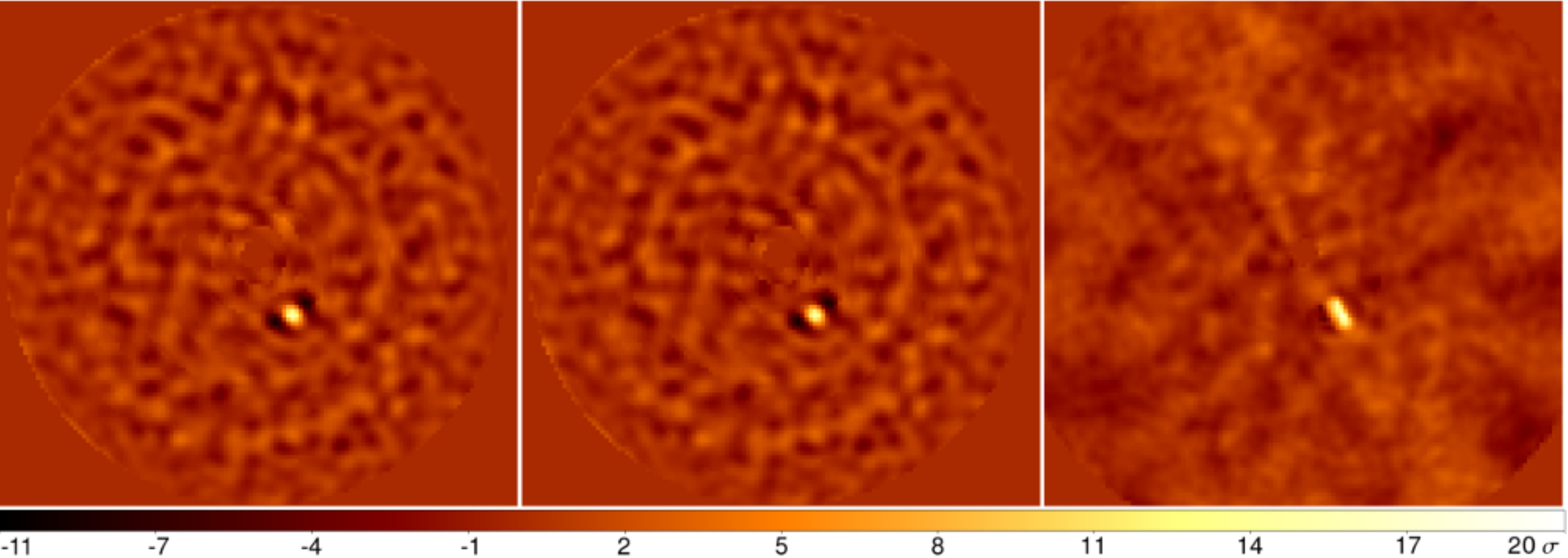}}
  \caption{Resulting S/N map of $\beta$ Pictoris according to two different image processing methods applied on the reduced data.
  \textit{Left:} S/N map obtained from ANDROMEDA, using a least-squares optimization for the ADI subtraction.
  \textit{Middle:} S/N map obtained from ANDROMEDA, using a L1-affine optimization for the ADI subtraction. \modif{The residuals are slightly lower, mostly close to the central part, as explained in Sect.~\ref{ADI}}.
  \textit{Right:} S/N map obtained with the PCA-KLIP algorithm (see text). All images are linearly scaled.
  \modif{No signal is found above the $5\sigma$ threshold in any of the maps, except for the companion $\beta$ Pictoris $b$.}}
\label{fig-BetaPic}
\end{figure*}

\begin{table*}[t]
\centering
\caption{Relative astrometry and photometry of $\beta$ Pictoris $b$ retrieved from the VLT/NaCo-AGPM data processed with PCA-KLIP and ANDROMEDA methods.}
\label{tab-BetaPicPCA}
\begin{tabular}{l c c}
\hline \hline
\textbf{Parameters} & \textbf{Estimated value using ANDROMEDA} &  \textbf{Estimated value using PCA} \\
\hline
\modif{Peak} S/N ($\sigma$):            & $\mmodif{17} $                                        & $\mmodif{17} $          \\
Separation from the star $sep$ (mas):   & $\mmodif{455   \pm 8.7}$                              & $452   \pm 9.6$         \\
Position angle $PA$ ($^\circ$):         & $\mmodif{210.7 \pm 0.8}$                              & $211.2 \pm 1.3$ \\
Contrast $\Delta L'$ (mag):             & $\mmodif{8.09  \pm 0.21}$                             & $8.01  \pm 0.16$        \\
\hline
\end{tabular}
\end{table*}

\subsection{ANDROMEDA results and comparison with PCA}
This section compares the quality of the detection, photometry, and astrometry obtained by ANDROMEDA with respect to the output of a principal component analysis \citep[PCA][]{Soummer2012,Amara2012}, as presented in \cite{Absil2013}. We thus ran ANDROMEDA on the exact same data cube and the automatic detection module on its output.

Given that for this study we have selected a region of $\mmodif{150 \times 150}$ pixels around the star, and knowing the pixel scale of the $L'$ camera ($\mmodif{27.15}$ mas/pixel), the user-defined parameter values chosen to run ANDROMEDA and the detection module are listed in Tab.~\ref{tab-BetaPicPara}.
\begin{table}[!h]
\centering
\caption{Parameters used to run ANDROMEDA on the $\beta$ Pictoris VLT-NaCo-AGPM data set.}
\label{tab-BetaPicPara}
\begin{tabular}{l c}
\hline \hline
\textbf{Parameter} & \textbf{Used value} \\
\hline
$F$                & 1/4 (default) \\
$dr$               & 1 $\lambda$/D (default)   \\
$R_A$              & 2 (default)   \\
$\delta_{\rm min}$ & 1 $\lambda$/D (default)  \\
$N_{psf}$          & $20 \times 20$ pixels  \\
$N_{smooth}$       & $\mmodif{10}$ pixels   \\
$IWA$              & 1 $\lambda$/D  \\
\hline
Threshold       & $5\sigma$ \\
Subwindow size  & $\mmodif{9} \times \mmodif{9}$ pixels \\
\hline
\end{tabular}
\end{table}

The resulting S/N map is shown in Fig.~\ref{fig-BetaPic}-\textit{Left} where the expected companion $\beta$ Pictoris $b$ is clearly visible as a sharp bright spot, southeast of the star (located at the center of the frame).

The same data set has been processed with a home-grown PCA algorithm, based on the KLIP implementation \citep{Soummer2012}. The KLIP algorithm uses a truncated basis of eigenvectors created by a Karhunen-Lo\`eve transform of the initial set of images, to perform the subtraction of the star residuals. \modif{To follow the original KLIP algorithm, the image processing was performed on the full $\mmodif{150\times150}$ pixel frames at once, and no frame was excluded from the ADI image library based on the amount of parallactic angle variation \citep[unlike in][]{Absil2013}. From the final image of the PCA processing, we compute a S/N map by testing resolution elements centered on each pixel of the map against all the other resolution elements located at the same angular distance from the star, as described in \citet{Mawet2014}. The number of principal components used in KLIP was tuned to maximize the S/N of the planet in the final image, resulting in the use of 18 principal components instead of 30 in \citet{Absil2013}. The S/N map is illustrated in Fig.~\ref{fig-BetaPic}-\textit{Right}, showing a peak S/N $\sim17$ on the  planet, i.e., the same as in the case of ANDROMEDA. We note that more advanced implementations of the KLIP algorithm, working on well-defined subregions (e.g., annuli or parts of annuli) in the images and including a frame rejection criterion based on the parallactic angle, can reach a peak S/N of up to $\sim20$, although at the expense of a drastically increased computation time (a few minutes instead of $\sim1$ second for the classical KLIP algorithm). Using ANDROMEDA, we could reach this S/N value by smoothing the profile over 20 pixels instead of ten, as used here. This does not affect the computation time, but for this value of $N_{smooth}$ one artifact appears above the $5\sigma$ threshold.}

The errors on the photometric and astrometric estimations are mostly limited by instrumental calibration errors, which can be decomposed into three main contributions: the error on the position of the source (8.5~mas), the PSF centering error in the images (0.1~mas), and the plate scale error (0.04~mas). As these errors are independent, the resulting error is the quadratic sum, giving in this case a total of 8.5~mas. Adding the $3\sigma$ error on the position estimation with ANDROMEDA (\modif{1.7~mas}), the total error on the position is of \modif{8.6~mas and the algorithm's intrinsic error is evaluated within this error bar}. For the position angle, the main errors are due to the true north direction knowledge ($0.09^\circ$) and the offset of the derotator ($0.01^\circ$), as discussed by \citet{Chauvin2012}. ANDROMEDA gives the PA with a $3\sigma$ precision of $\mmodif{0.8^\circ}$ and a negligible intrinsic error. The errors on the flux estimation are mainly due to the variation of the PSF along the observation (0.05~mag). ANDROMEDA estimates the flux with a $3\sigma$ statistical error of \modif{0.2~mag} and a negligible intrinsic error, and as these three sources are independent again, the total error is the quadratic sum of each error. The estimated position and contrast with both methods and their corresponding error bars are found in Tab.~\ref{tab-BetaPicPCA}, where the photometric and astrometric estimations for the PCA pipeline are directly taken from the publication of \cite{Absil2013}. The two estimations agree with each other within error bars. 

\modif{We emphasize once again here that the photometry and astrometry can be recovered much more easily with ANDROMEDA than with PCA, in particular, because the estimation is direct and precise without needing to inject fake companions. The same  applies for the detection limit computation, which is  given directly by the map of the standard deviation of the flux provided by ANDROMEDA and does not need, for instance, numerous synthetic planet injections as for other methods.}

\begin{figure*}[t]
\centering
\resizebox{\hsize}{!}{\includegraphics[scale=0.05]{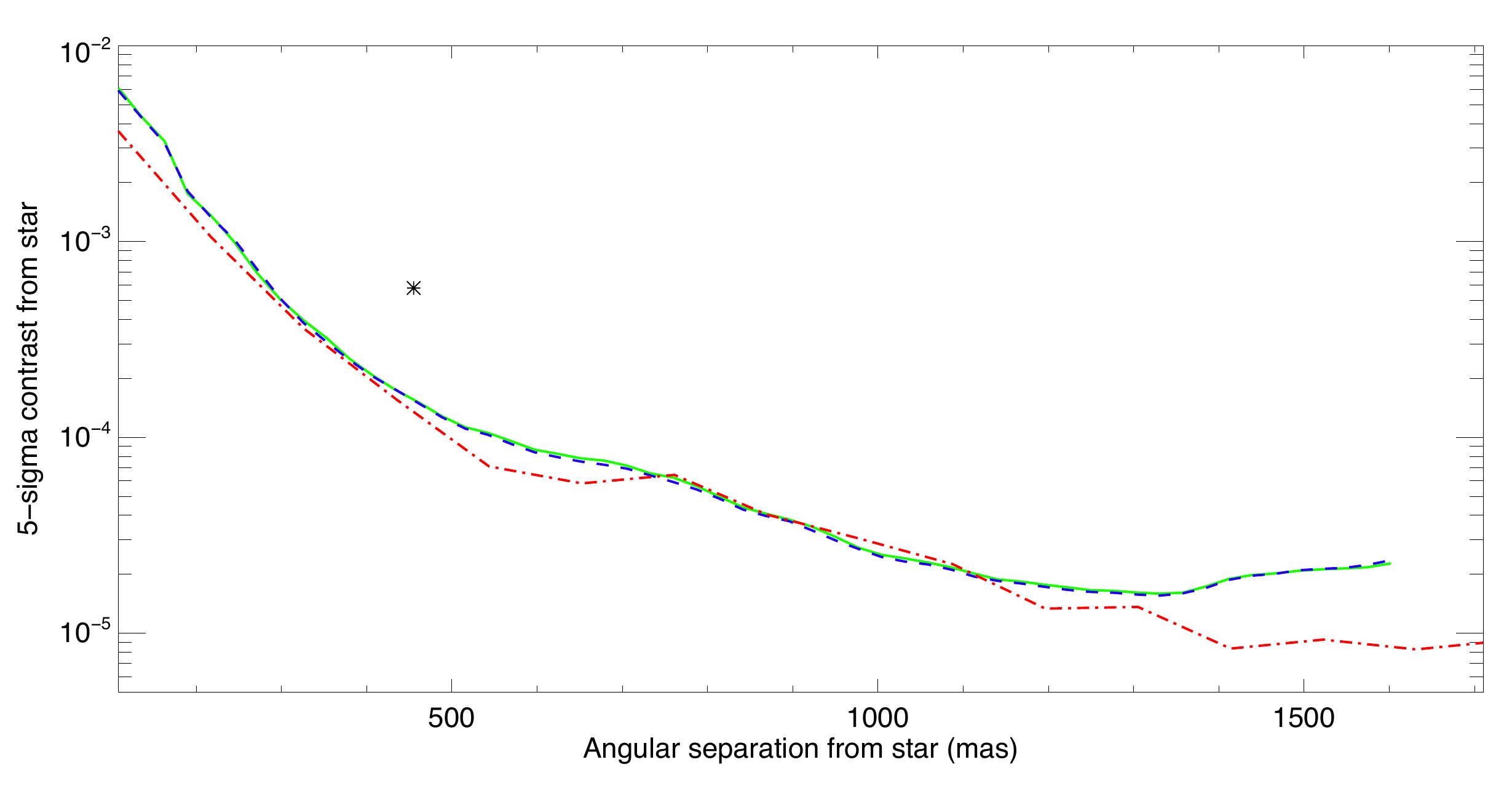} }
\caption{Detection limit\modif{s at $5\sigma$} obtained by processing the $\beta$ Pictoris data from VLT/NaCO-AGPM with ANDROMEDA \modif{(green solid line, when using a L1-affine optimization or blue dashed line when using a LS optimization)} or \modif{sPCA} \modif{(red dash-dotted line)}. The asterisk symbol shows the location of the detected signal above the $5\sigma$ threshold set (only $\beta$ Pictoris b is detected here).}
\label{fig-BetaPicDetLim}
\end{figure*}

\subsection{Sensitivity comparison and discussion}
\label{sec-comparison}
Another way of comparing the performance of the two methods is to plot their $5\sigma$ detectability limits in terms of contrast.

For the PCA-KLIP method, \modif{the computation of the detection limits is based on} the hypothesis that the noise is Gaussian \citep{Marois2008}, which was checked empirically in \cite{Absil2013}. Several methods have been used in the literature to \modif{derive detection limits}. Here, we use the statistical framework presented in \citet{Mawet2014}, where test speckles are compared to the statistics of the flux within all other resolution elements located at the same angular distance from the center. This framework assumes that the noise statistics varies with the distance from the star but not with the azimuth, and it takes into account the effect of small sample statistics on the noise estimations. More specifically, the computation of the contrast curve consists in normalizing the flux in the final PCA-processed image by the integrated flux contained inside the (off-axis) stellar PSF core, then in calculating the standard deviation of the fluxes enclosed in all the non-overlapping apertures of diameter equal to the FWHM of the PSF that can be placed at a given distance from the center. The same procedure is repeated at all radial distances. During the process, any known companion can be masked out , which we did here in order not to bias our noise estimation at the distance of $\beta$~Pic~b. This estimation of the contrast curve provides more realistic results than the pixel-to-pixel standard deviation noise estimation that has frequently been used in the literature in the past.

\modif{We chose to use here a ``smart'' version of the PCA-KLIP algorithm (sPCA), where the images are decomposed in annuli and where a frame rejection criterion is included based on the parallactic angle, in order to avoid self-subtraction of the possible planetary signals. We used the same smart PCA parameters as for the computation of the detection limits in \citet{Absil2013}.} The final contrast curve, plotted in Fig.~\ref{fig-BetaPicDetLim}, takes into account the algorithm throughput, estimated at several radial distances (and azimuths) by injecting fake companions in the original data cube, running the PCA pipeline with the exact same parameters, and comparing the flux of the fake companions in the final PCA-processed image with the flux of the injected companions. We note that the PCA contrast curve presented here is a factor of two higher than the one presented in \citet{Absil2013}, where an improper convolution by a Gaussian kernel led to a factor of two underestimation of the contrast curve.

Figure~\ref{fig-BetaPicDetLim} shows the different detection limit curves obtained \modif{with the sPCA algorithm and with} ANDROMEDA using the methods noted above, all corrected for the small sample statistics. The location of $\beta$ Pictoris $b$ in terms of angular separation and contrast is also indicated on the graph for comparison. In this figure, we can see that at angular separations ranging from 0.1\arcsec to 2\arcsec, ANDROMEDA and PCA provide similar detection limits.

\modif{To give a rough idea, we mention here that ANDROMEDA takes about 180 seconds to process the 612 $150\times150$ pixels images on a quadri-pro six-core Intel Xeon. This computing time includes the time to obtain the three main outputs from ANDROMEDA (estimated flux map, S/N map, and the map of the standard deviation of the estimated flux) plus the detection  described in Sect.~\ref{sect-Toolbox} (detection of probable point source, sorting out, subpixel position estimation, flux estimation, and corresponding derivation of the  $3 \sigma$ error bars  and the computation of the detection limit curve). This latter detection  is obtained from the ANDROMEDA output within a few seconds. For the PCA-KLIP method, it takes about 1 second to produce the final reduced image using the original KLIP algorithm. Building the S/N map in Fig.~\ref{fig-BetaPic} from this reduced image along the \citet{Mawet2014} prescription takes an additional few tens of seconds. We note that the annulus-wise, ``smart'' PCA (sPCA) processing used to compute the detection limits in Sect.~\ref{sec-comparison} takes about 10 minutes on a recent core-i7 laptop. Also, we emphasize here that the photometry and astrometry are included in ANDROMEDA while PCA needs dedicated, computationally intensive methods to get an accurate estimation (such as the negative fake companion technique, which requires running many PCAs to obtain the photometric and astrometric estimations together with their error bars). The latter figures must be taken with care since there are still ways to optimize the computation time for both algorithms (for instance parallelizing the ADI plus MLE with ANDROMEDA), but they are mentioned here to give a rough order of magnitude of the processing time.}

The two algorithms that we have used to process the VLT/NaCo-AGPM data of $\beta$ Pictoris both efficiently detect the close companion $\beta$ Pictoris $b$. However, ANDROMEDA does the detection automatically, by giving  direct access to the S/N map (which is only a by-product in the PCA algorithm). The approaches that the two algorithms use to estimate the position and flux of the companion are quite different, but result in compatible values. ANDROMEDA potentially shows better accuracy in terms of separation and contrast estimation since it directly relies on the detection probability derived from a strong hypothesis about the image formation and the residual noise in the images, which are well verified. To conclude, the two methods are complementary since they do not rely on the same inner concept and it is better to use several methods that are significantly different to better judge of the truthfulness of a detection.


\section{Conclusion and perspectives}
The ANDROMEDA method is designed to detect planetary-like signals in high-resolution and high-contrast images. 
In this paper we have described the improvements brought to the original method when applied to on-sky VLT/NaCo data, in order to make it more robust.
This included the implementation of a pre-processing (high-pass filtering of the data), a post-processing (normalization of the S/N map), and the use of a robust method to perform the image difference. As a result, this method proves to be highly efficient at processing experimental data from the VLT/NaCo instrument taken in pupil-tracking mode.
In particular the produced normalized S/N map can be thresholded with a constant value throughout the field of view so as to provide detectability and false alarm levels that are homogeneous over the field. This allows one to automatically detect  the candidate companions present in the image, and classify them  according to their S/N values. Artifacts can then be rejected based on morphogical criteria expected for a true companion. We then showed on synthetic companions and on the well-known case of $\beta$ Pic$\,$b that ANDROMEDA accurately estimates the subpixel position and the contrast of these companions. Notably, and unlike most  other methods, the selection of the data reduction parameters (like filtering or minimum angular separation used for image differences) is fully taken into account in the search for the companion signature, and thus it does not bias the companion flux estimate. In particular, the analysis of the result does not require subsequent tests with synthetic planet injections to estimate a companion flux loss.

The second generation of instruments dedicated to exoplanetary detection and characterization, such as VLT-SPHERE, \citep{Beuzit08}, Gemini-GPI, \citep{Macintosh2008}, and Subaru-SCeXAO, \citep{Guyon2009,Guyon2013}, \modif{is about to} deliver a large amount of data that will require massive, homogenous, and efficient companion extraction. The capabilities of ANDROMEDA in terms of performance and automatic detection will be very beneficial in this context.

Though ANDROMEDA is operational in its current state, any further experience on these new instruments may motivate the evolution of this algorithm\footnote{The authors of the ANDROMEDA code are open to working in collaboration in order to apply the code to other data and thus gain further experience.}. In particular, its probabilistic formalism directly enables one to include any \textit{\emph{a priori}} knowledge of the noise structure of the data or on the sought companions. Evolutions can be considered in various ways to build \modif{differential images} (for instance, if it appears that the closest images in time are not necessarily the most correlated). Finally, the non-coronagraphic PSF, defining the companion's expected signature, is currently assumed to be stable along the data sequence, as measured before or after the coronagraphic or saturated data set. If the evolution of the PSF and in particular the evolution of the Strehl ratio are known, this variability can be easily included in the algorithm for a better flux estimate fidelity.


\begin{acknowledgements}
The authors thank A.~ Eggenberger for preliminary work made on applying ANDROMEDA to real VLT/NaCo data.
We  would also like to especially thank D.~Mawet, R.~Galicher, C.~Lefevre, \modif{A.~Vigan} and T.~Fusco for their careful reading and relevant comments.
F.~Cantalloube and L.~Mugnier thank the European Commission under FP7 Grant Agreement No. 312430 Optical Infrared Coordination Network for Astronomy and  the French Aerospace Lab (ONERA) (in the framework of the NAIADE Research Project) for  partly funding this work.
O.~Absil \modif{and C.~A.~Gomez Gonzalez acknowledge} funding from the European Research Council Under the European Union's Seventh Framework Programme (ERC Grant Agreement n. 337569) and from the French Community of Belgium through an ARC grant for Concerted Research Action.
\end{acknowledgements}

%

\bibliographystyle{aa}
\bibliography{Andromeda}

\end{document}